\documentclass[notitlepage,showpacs,amsmath,aps,pra,amssymb,nofootinbib]{revtex4-1} 
\usepackage[T1]{fontenc}
\usepackage[latin9]{inputenc}
\usepackage{times}
\usepackage{color} 
\usepackage{xspace}
\usepackage{amssymb,amsmath}
\usepackage{amsbsy}
\usepackage{graphicx}
\usepackage{bm}
\usepackage{epstopdf}
\usepackage{float}
\usepackage{booktabs}
\usepackage{adjustbox}

\usepackage[unicode,breaklinks]{hyperref}
\hypersetup{
    unicode=true,
    a4paper=true,
    plainpages=false, 
    colorlinks=true,
    linkcolor=blue,
    citecolor=blue,
    filecolor=black,
    urlcolor=blue
}
\newcommand{\bk}{\mathbf{k}}

\newcommand{\br}{\mathbf{r}}
\newcommand{\bx}{\mathbf{x}}

\newcommand{\brho}{{\bm{\rho}}}

\newcommand{\nt}{{\tilde{n}}}
\newcommand{\mub}{\bar{\mu}}
\newcommand{\Ebz}{\bar{E}_z}
\newcommand{\hz}{\bar{h}_z}

\newcommand{\UD}{U_{\mathrm{dd}}}

\newcommand{\Ut}{\tilde{U}}
\newcommand{\chib}{\bar{\chi}}

\begin{document}
\begin{samepage}

\title{A general theory of flattened dipolar condensates} 
\author{D.~Baillie}
\author{P.~B.~Blakie}  
\affiliation{Dodd-Walls Centre for Photonic and Quantum Technologies, Department of Physics, University of Otago, Dunedin, New Zealand}

\begin{abstract} 
We develop theory for a flattened dipolar Bose-Einstein condensate (BEC) produced by harmonic confinement along one direction. The role of both short-ranged contact interactions and long-ranged dipole-dipole interactions (DDIs) is considered, and the dipoles are allowed to be polarised along an arbitrary direction. We discuss the symmetry properties of the condensate and the part of the excitation spectrum  determining stability, and introduce two effective interaction parameters that allow us to provide a general description of the condensate properties, rotons, and stability.  We  diagonalize the full theory to obtain benchmark results for the condensate and quasiparticle excitations, and characterize the exact mean field stability of the system. We provide a unified formulation for a number of approximate schemes to describe the condensate and quasiparticles, including the standard quasi-two-dimensional (quasi-2D) approximation, two kinds of variational ansatz, and a Thomas-Fermi (TF) approximation. Some of these schemes have been widely used in the literature despite not being substantiated against the exact theory. We provide this validation and establish the regimes where the various theories perform well. \end{abstract}

\maketitle

%==============================================================================
 \section{Introduction}
Dipolar Bose-Einstein condensates (BECs) have been produced with chromium  \cite{Griesmaier2005a,Beaufils2008a}, dysprosium \cite{Mingwu2011a}, and erbium \cite{Aikawa2012a} atoms. All of these atoms have a large magnetic moment, causing them to interact with  an appreciable dipole-dipole interaction (DDI), which is both long-ranged and anisotropic \cite{Baranov2008,Lahaye_RepProgPhys_2009}. There is significant interest in flattened dipolar BECs, produced by an external trapping potential that confines one direction tightly. For example, experiments have confirmed that flattened dipolar BECs, with dipoles polarised normal to the trap plane, are stable against an attractive short-ranged contact interaction \cite{Koch2008a,Muller2011a}. Theoretical predictions for this regime include the emergence of novel density oscillating condensate ground states \cite{Ronen2007a,Lu2010a,Asad-uz-Zaman2010a,Martin2012a}, enhanced density fluctuations \cite{Bisset2013a,Boudjemaa2013a,Blakie2013a},  roton-like excitations  \cite{Santos2003a,Ronen2007a,Nath2010a,Hufnagl2011a,Blakie2012a,JonaLasinio2013,Bisset2013b}, and modified collective and superfluid properties \cite{Wilson2010a}. Additionally, for a negative DDI  (e.g.~produced by rapidly rotating the dipoles \cite{Giovanazzi2002a}) it has been predicted that stable 2D bright solitons can be produced \cite{Pedri2005a}. We also note that stacks of flattened systems (e.g.~as can be produced by an optical lattice) are of interest (e.g.~see \cite{Klawunn2009a,Junginger2010a,Muller2011a}), where inter-layer interactions add to the richness of the physics. Tilting the dipoles relative to the flattened trap leads to anisotropy in superfluid properties, spectra and density fluctuations \cite{Ticknor2011a,Bismut2012a,Baillie2014a}.
\end{samepage}

Quantitative calculations for the mean field ground states and quasiparticle excitations of flattened dipolar BECs are difficult, because fundamentally the problem is still fully three-dimensional (3D) and requires a number of specialised numerical techniques to deal with the DDI \cite{Ronen2006a}. As such the most comprehensive calculations  are restricted to cylindrically symmetric regimes, which requires the dipoles to be polarised along the symmetry axis of the trap.
 One simplification, commonly referred to as the \textit{quasi-2D approximation} \cite{Fischer2006a,Cai2010a}, is based on the assumption that the system is in the harmonic oscillator ground state in the tightly confined direction, which can then be integrated out leaving an effective 2D theory. This approximation is only valid when interactions are weak, and many regimes of interest (e.g.~roton excitations or the collapse instability) occur where this condition is violated \cite{Fischer2006a} (also see \cite{Parker2008a}).
 For systems with 3D confinement, a comparison of the quasi-2D approximation to other approaches was presented by Wilson \textit{et al.}~\cite{Wilson2011a}. That work suggests that the quasi-2D approximation furnishes a qualitatively useful description, even though it is quantitatively inaccurate, e.g.~the DDI strength at which collapse occurs is overestimated by a factor of about three (see Fig.~3 of \cite{Wilson2011a}). 
An increasing number of studies of flattened dipolar BECs have been based on the quasi-2D approximation (e.g.~see \cite{Cai2010a,Wilson2012a,Natu2013a,Natu2013b,Corson2013b,Mulkerin2013a}). 
It is therefore essential to develop a better quantitative theory for the flattened system in the regime where the quasi-2D approximation is only qualitatively correct, particularly because of the interesting physics that is predicted to occur in this regime.
\begin{figure}[ht!] 
   \centering 
   \includegraphics[width=3.5in]{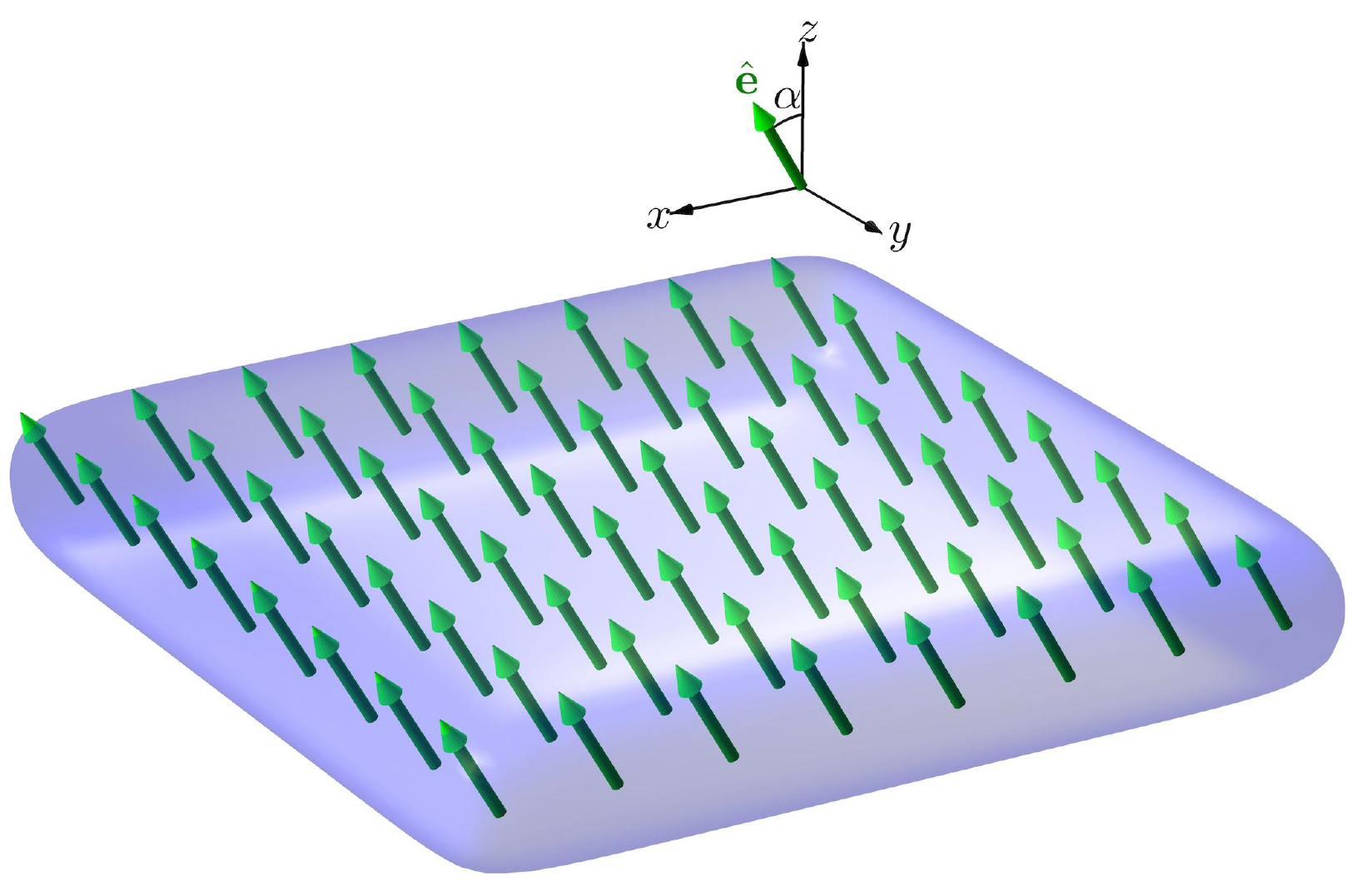}
  \caption{Schematic of  a planar dipolar condensate harmonically confined along $z$ with dipoles polarised in the $xz$-plane at an angle of $\alpha$ to the $z$-axis.
   \label{f:schematic} }    
\end{figure}

 In this paper we present a general treatment of a flattened dipolar BEC. We focus on the situation where the confinement is harmonic along one direction, with free in-planar motion. We allow for the dipoles to be tilted at an arbitrary angle into the plane. We formulate mean field equations for the condensate and quasiparticle excitations. We observe that the condensate possesses a symmetry, whereby its properties are determined by a single effective interaction parameter accounting for the contact  interaction, DDI, and the effect of tilt.
 Additionally, we show that the part of the excitation spectrum that determines the onset of rotons and collapse instability possesses another rigorous symmetry. This motivates the introduction of a second effective interaction parameter, and allows us to provide  a universal description of the stability boundary. We numerically solve the exact mean field equations for condensate and quasiparticles. This allows us to obtain benchmark results for the universal properties of the condensate and the stability phase diagram, which are provided as data files with this paper.\footnote{The data from the full numerical calculations is included with this arXiv submission.} We then turn to considering approximate descriptions of the system which involve a non-trivial reduction of the system to a 2D problem by assigning a specified profile for the condensate and quasiparticle profiles in the  confined direction. In detail we consider the standard quasi-2D approximation, a  Gaussian (i.e.~\textit{single mode}) and two-mode variational ansatz, and a Thomas-Fermi (TF) approximation. We unify the formulation of these approximate descriptions and consider how well they perform relative to our exact results at predicting the properties of the condensate, the excitation spectrum and stability.

\section{ Formalism}\label{s:formalism}
\subsection{Gross-Pitaevskii equation}\label{S:GPE}
We consider here a dipolar BEC that is harmonically confined along the $z$ direction and unconfined in the radial plane. 
The condensate wave function $\psi_0$ satisfies the non-local Gross-Pitaevskii equation  (GPE)
\begin{equation}
\mu\psi_0=\left[-\frac{\hbar^2\nabla^2}{2m}+\frac{m\omega_z^2z^2}{2}+\int d\br'U(\br-\br')|\psi_0(\br')|^2\right]\psi_0,\label{e:fullGPE}
\end{equation}
where $\omega_z$ is the trap frequency and $\mu$ is the chemical potential. We consider DDIs for the case of dipoles polarised by an external field along the direction $\hat{\mathbf{e}}=\hat{\mathbf{z}}\cos\alpha+\hat{\mathbf{x}}\sin\alpha$, lying in the $xz$-plane. We refer to $\alpha$ as the tilt angle of the dipoles, with the untilted case $\alpha=0$ corresponding to dipoles polarised along $z$. The associated interaction potential is 
\begin{equation}
\UD(\br)=\frac{3g_d}{4\pi}\frac{1-3(\hat{\mathbf{e}}\cdot\hat{\mathbf{r}})^2}{r^3},\label{Eqn:DDI}
\end{equation}
where $g_d=\mu_0\mu_m^2/3$ is the DDI coupling constant and $\hat{\mathbf{r}}=\mathbf{r}/|\mathbf{r}|$. The particles can also interact by a short ranged contact interaction with coupling constant $g_s=4\pi a_s\hbar^2/m$, where $a_s$ is the scattering length, so that the full interaction we use is $U(\mathbf{r})=g_s\delta(\br)+U_{\mathrm{dd}}(\br)$ \cite{Yi2000a,Yi2001a}.

The condensate solution takes the form $\psi_0(\br)=\sqrt{n}\chi_\nu(z)$, with $n$ the areal density, and the normalised axial mode $\chi_\nu$ satisfies (see Appendix \ref{s:intpotGPE})
\begin{align}
\mu\chi_\nu(z)&= \mathcal{L}_{\mathrm{GP}}\chi_\nu(z),\label{e:GPE1d}\\
\mathcal{L}_{\mathrm{GP}} &\equiv \hbar\omega_z[ \hz+\nu l_z|\chi_\nu(z)|^2],\label{e:LGPE}\\
\hz &\equiv \frac{1}{\hbar\omega_z}\left(-\frac{\hbar^2}{2m}\frac{d^2}{dz^2}+\frac{m\omega_z^2z^2}{2}\right),\\
\nu &\equiv\frac{n[ g_s+g_d(3\cos^2\alpha-1)]}{\hbar\omega_zl_z},\label{e:nu}
\end{align}
where $l_z=\sqrt{\hbar/m\omega_z}$ and from Eq.~\eqref{e:GPE1d} we see that the GPE solution depends only\footnote{i.e.~in oscillator units  the dimensionless transverse mode  $\chib_\nu(u)\equiv\chi_\nu(ul_z)\sqrt{l_z}$ only depends on $\nu$} on the value of the dimensionless effective interaction parameter $\nu$. 
Thus, while the interaction $U(\mathbf{r})$ is characterised by the  parameters $\{g_s,g_d,\alpha\}$, the GPE solution is invariant on surfaces in $\{g_s,g_d,\alpha\}$-space where the value of $\nu$ is constant.

\begin{figure}[ht!] 
   \centering 
   \includegraphics[width=3.5in]{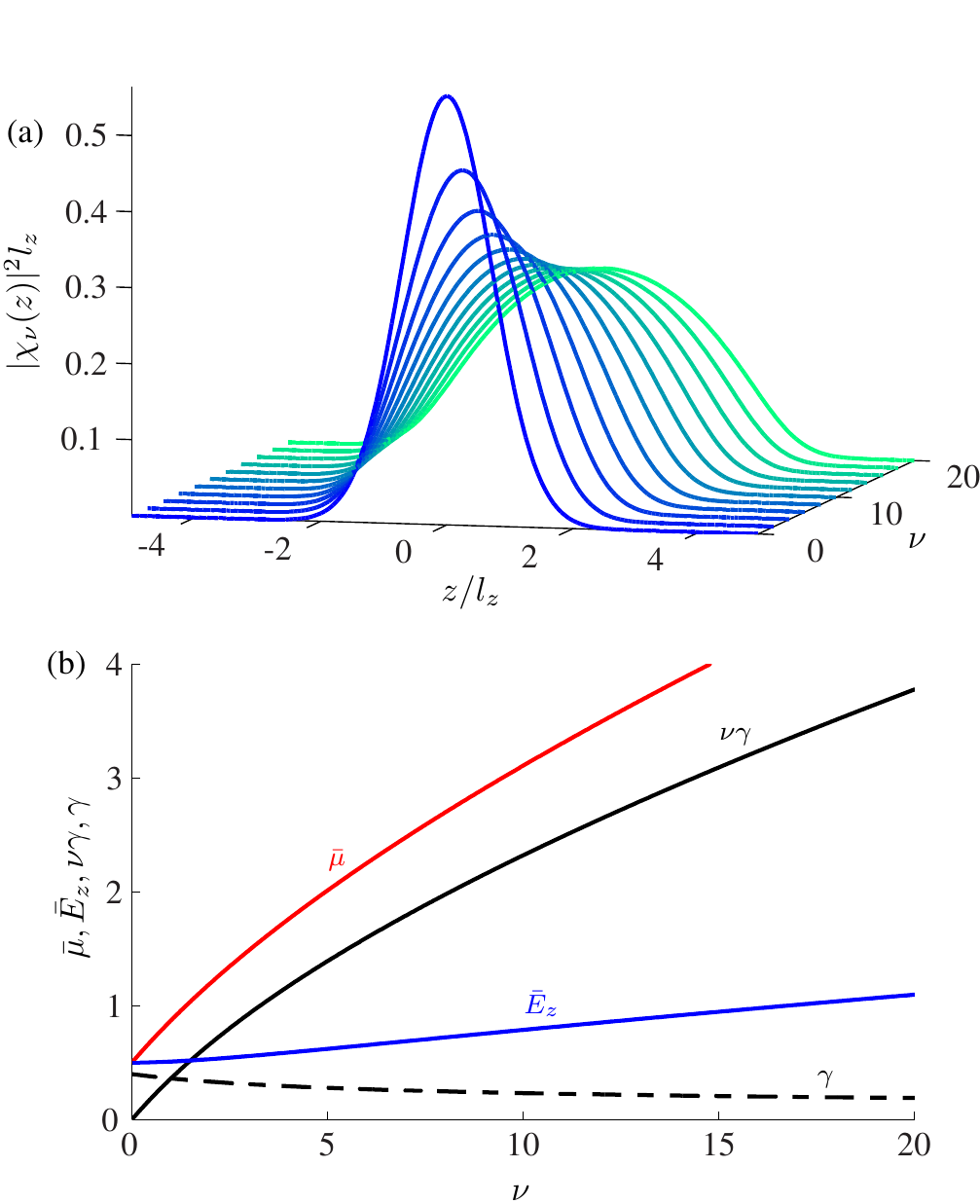}
  \caption{(a) Condensate mode $\chi_\nu$ obtained by numerical solution of the GPE \eqref{e:GPE1d} for $\nu$ from $0$  to $20$ in steps of $2$. (b) The dimensionless parameters $\bar{\mu}(\nu)$, $\Ebz(\nu)$, $\nu\gamma(\nu)$, and $\gamma(\nu)$.
   \label{f:chis} }    
\end{figure}

We perform numerical calculations to solve Eq.~(\ref{e:GPE1d}) for the condensate mode $\chi_\nu$, by discretizing it at a set of evenly spaced grid points along $z$. The discretized GPE is cast in a normalisation insensitive form \cite{Modugno2002a}, and then solved using a Newton-Krylov solver \cite{KellyBook}. In Fig.~\ref{f:chis}(a) we show results for $\chi_\nu$ for a range of values of $\nu$. The $\nu=0$ solution is the ideal harmonic oscillator state, and as the value  of $\nu$ increases the solution broadens out in the trap to reduce the interaction energy. A variety of approximate solutions for the condensate are discussed in Sec.~\ref{s:ApproxTreatments}.

An important characterisation of the GPE solution is provided by the quantities
\begin{align}
\bar{\mu}(\nu)&\equiv\frac{\mu}{\hbar\omega_z},\label{q:mubar}\\
\gamma(\nu)&\equiv\int dz\,l_z|\chi_\nu(z)|^4 \label{e:gamma},\\
\Ebz(\nu)&\equiv\int dz\,\chi_\nu^*(z)\hz  \chi_\nu(z), \label{e:Ez}
\end{align}
i.e.~the (dimensionless) chemical potential, confinement parameter, and $z$-confinement energy, respectively. These quantities are functions of $\nu$ only, and are related to each other by $\mub(\nu) =  \nu\gamma(\nu) + \Ebz(\nu)$,  obtained by taking the expectation of $\mathcal{L}_{\mathrm{GP}}$. 
Their values, obtained from our numerical calculations for $\chi_\nu$, are plotted in Fig.~\ref{f:chis}(b) and are available as raw data files (see footnote 1). Noting that for the non-interacting case $\nu=0$, we have $\bar{\mu}=\bar{E}_z=\frac{1}{2}$ and $\gamma=1/\sqrt{2\pi}$.

\subsection{Bogoliubov excitations}
The excitations of a condensate are Bogoliubov quasiparticles. For our geometry these can be obtained by linearising the time-dependent GPE\footnote{Eq.~\eqref{e:fullGPE} with $\mu\to i\hbar\frac{\partial}{\partial t}$. The backaction of quasiparticles could be included in more advanced theories \cite{Griffin1996a}.} about the condensate using an ansatz (e.g.~see \cite{Morgan1998a,Ronen2006a}) of the form 
\begin{align}
&\psi(\mathbf{r},t)=e^{-i\mu t}\left\{\sqrt{n}\chi_\nu(z)\, + \sum_{\bk_\rho,j}\left[c^j_{\bk_\rho}e^{-iE^j_{\bk_\rho}t/\hbar}u^j_{\bk_\rho}(z)e^{i\bk_\rho\cdot\brho}-c^{j*}_{\bk_\rho}e^{iE^j_{\bk_\rho}t/\hbar}v^{j*}_{\bk_\rho}(z)e^{-i\bk_\rho\cdot\brho}\right]\right\},\label{EqnBog}
\end{align}
where the $\{c^j_{\bk_\rho}\}$ are (small) c-number amplitudes, and the $\{u^j_{\bk_\rho}(z),v^j_{\bk_\rho}(z)\}$ are quasiparticle amplitudes, with respective eigenenergies $\{E^j_{\bk_\rho}\}$. The amplitudes  satisfy the normalisation condition $\int dz ({|u^j_{\bk_\rho}|^2-|v^j_{\bk_\rho}|^2})=1$. In the above expansion we have introduced the planar coordinate vector $\bm{\rho}=(x,y)$ and wave vector $\bk_\rho=(k_x,k_y)$. Because the system is homogeneous in the $\bm{\rho}$-plane, the quasiparticles are of well defined momenta $\hbar\bk_\rho$. The additional index $j$ labels the $z$ vibrational mode.

The Bogoliubov-de Gennes (BdG) equations that must be solved to determine $\{E_{\bk_\rho}^j,u_{\bk_\rho}^j(z),v_{\bk_\rho}^j(z)\}$  take the form (see Appendix~\ref{s:BogEqns}) 
\begin{equation}
E_{\bk_\rho}^j   \begin{bmatrix} u_{\bk_\rho}^j(z)\\ v_{\bk_\rho}^j(z)  \end{bmatrix}
    =   \begin{bmatrix}       \mathcal{L} & -\mathcal{M}\\
             \mathcal{M}^*  & -\mathcal{L}^* 
   \end{bmatrix}
     \begin{bmatrix} 
      u_{\bk_\rho}^j(z)\\
     v_{\bk_\rho}^j(z)
 \end{bmatrix},\label{e:BdG}
\end{equation}
where
\begin{align}
    \mathcal{L}f^j_{\bk_\rho}(z)&= \left(\mathcal{L}_{\mathrm{GP}}+\frac{\hbar^2k_\rho^2}{2m}-\mu+\mathcal{M}\right)f^j_{\bk_\rho}(z),\\
    \mathcal{M}f^j_{\bk_\rho}(z) &= n\chi_\nu(z)\mathcal{F}^{-1}\left\{ \Ut(\bk) \mathcal{F}\left\{ \chi_\nu(z) f^j_{\bk_\rho}(z)\right\}\right\},\label{e:Mf}\\
    \Ut(\bk) &= g_s+g_d\left[\frac{3(k_z\cos\alpha  +k_x\sin\alpha )^2}{k^2}-1\right],\label{Eqn:DDIk}
 \end{align}
 $\mathcal{F}\{f\} = \int dz\,e^{-ik_zz} f$ denotes the Fourier transform in $z$, $\mathcal{F}^{-1}$ denotes its inverse, and \eqref{Eqn:DDIk} is the Fourier transform in $\br$ of the interaction $U(\br)$.

\begin{figure}[ht!] 
   \centering 
   \includegraphics[width=3.5in]{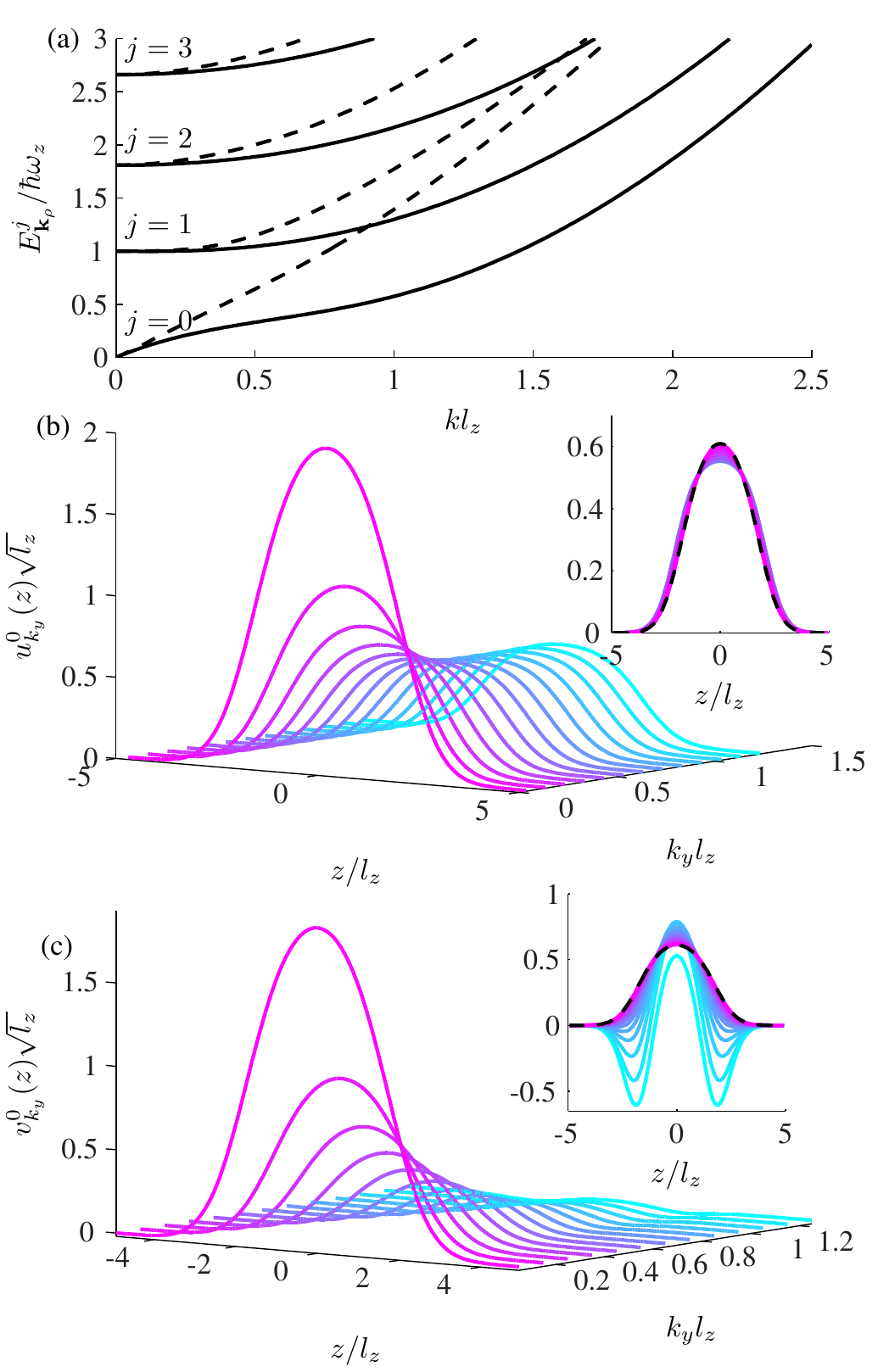}
  \caption{(a) Spectrum along any direction for $\alpha=0$ (solid, also along $k_y$ axis for any $\alpha$) and along $k_x$ axis for $\alpha=\pi/3$ (dashed), (b) $u_{k_y}^0$ and (c) $v_{k_y}^0$  amplitudes as a function of $k_y$ along $k_y$ axis, for $\nu=5$, $\nu_d=2$. Inset to (b) shows the $u_{k_y}^0$ (same colour lines as main plot) modes after they are normalised to reveal shape changes, with $\chi_\nu$ shown for reference (dashed). Similarly for the inset to (c).
   \label{f:euvs} }
\end{figure}

We numerically solve the BdG equations (\ref{e:BdG}) using the same discretization we employed to solve for $\chi_\nu$. The discretized equations are then diagonalised using standard linear algebra packages \cite{Garbow1974a}. For accurate calculation of the excitations it is important to employ a cutoff interaction, see Appendix~\ref{s:CutoffInt} (also \cite{Ronen2006a}).  An example of the spectrum and ground band quasiparticle amplitudes is shown in Fig.~\ref{f:euvs}. We note that the ground band  spectrum in Fig.~\ref{f:euvs}(a) shows some key characteristics of a planar dipolar condensate: phonon (linear) behaviour at low momentum (up to $k\sim0.3/l_z$), a softened spectrum at $k\sim1/l_z$ arising from the attractiveness of the DDI, and free particle (quadratic) behaviour at high $k$ where the kinetic energy dominates. For stronger dipolar interactions the spectrum can develop a roton minimum at $k\sim1/l_z$. The role of interactions in higher bands ($j>0$) is less important [see Fig.~\ref{f:euvs}(a)].

\subsection{$k_y$-spectrum symmetry}
 In general the quasiparticles must be solved numerically, however the spectrum along the $k_y$ axis   [i.e.~$E_{\bk_\rho}^j$ for $\bk_\rho=(0,k_y)$], possesses an interesting symmetry property.\footnote{The $y$-direction is special because $\hat{\mathbf{e}}$ is in the $xz$-plane.} To see this we note that for excitations along $k_y$, which we denote by setting $\bk_\rho\to k_y$ in the subscripts, Eq.~\eqref{e:Mf} simplifies to
\begin{align}
         &\frac{\mathcal{M}[f^j_{k_y}(z)]}{\hbar\omega_zl_z} = \nu f^j_{k_y}(z)[\chi_\nu(z)]^2
         -3\nu_d k_y^2\chi_\nu(z)\mathcal{F}^{-1}\left\{\frac{\mathcal{F}\left\{\chi_\nu(z)f^j_{k_y}(z)\right\}}{k_y^2+k_z^2}\right\}, \label{e:Mfky}
     \end{align}
using \eqref{Eqn:DDIk}, where we have defined
\begin{equation}
\nu_d \equiv \frac{ng_d\cos^2\alpha}{\hbar\omega_z l_z}.\label{e:nud}
\end{equation}
The overall action of $\mathcal{M}$ remains constant if $\nu$ and $\nu_d$ remain constant. These two constraints define a curve in $\{g_s,g_d,\alpha\}$ parameter space where the $k_y$ spectrum is invariant (c.f.~the GPE solution $\chi_\nu(z)$ which is invariant on $\nu$ constant surfaces). E.g., the $k_y$-spectra for the parameter sets $\{g_s,g_d,\alpha\}$ and $\{g_s',g_d',\alpha'\}$ are identical if
\begin{subequations}
\begin{align}
    g_s'&=g_s +g_d'- g_d = g_s+g_d\left(\frac{\cos^2\alpha}{\cos^2\alpha'}-1\right),\label{e:map1}\\
g_d'&=g_d\frac{\cos^2\alpha}{\cos^2\alpha'}.\label{e:map2}
\end{align}
\end{subequations}

This result is useful because for $g_d>0$ the stability of the system, and the development of roton excitations, is addressed by examining excitations on the $k_y$-axis, and in the $j=0$ band. We note that if $\nu_d<\nu/3$ then the interaction along $k_y$ is always repulsive.

\begin{figure}[ht!] 
   \centering 
   \includegraphics[width=3.5in]{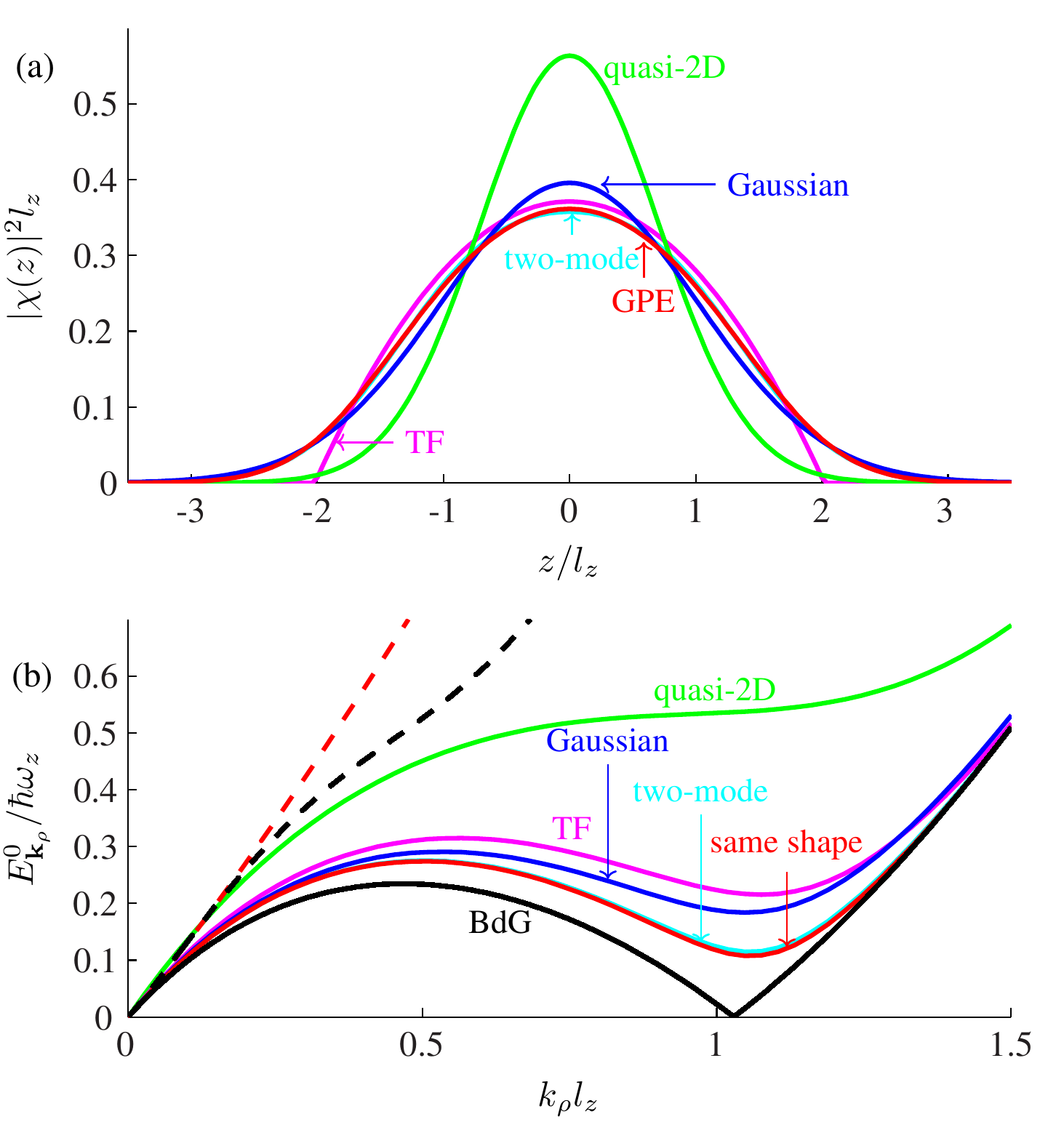}
  \caption{(a) Condensate mode $\chi$ obtained by numerical solution of the GPE \eqref{e:GPE1d}, with approximate results based on the variational, TF and quasi-2D approximations.
  (b) Quasiparticle spectrum for $j=0$ obtained by (black) full numerical solution of the BdG equations~\eqref{e:BdG}, with approximate results based on the same shape,  variational  and  quasi-2D approximations. 
Parameters:   $g_s=\alpha=0$ and $\nu_d=2.7434$ with results independent of direction in the $\bk_\rho$ plane.  This solution also applies to  $\alpha'=\frac{\pi}{3}$ using the mapping \eqref{e:map1}-\eqref{e:map2} with spectrum along the $k_y$ axis (solid, same as $\alpha=0$) and along the $k_x$ axis (dashed, only shown for BdG and the same shape approximation).
   \label{f:chispec} }
\end{figure}

In Fig.~\ref{f:chispec} we show the condensate mode and excitation spectrum for a dipolar system with $g_s=\alpha=0$ and $\nu_d=2.7434$ (i.e.~$\nu=5.4868$). 
 The spectrum reveals that the system has a roton (local minima) at $k_\rho\approx1/l_z$, and with approximately zero energy. Increasing $g_d$ further will cause the roton quasiparticles to become dynamically unstable, leading to local collapse of the condensate (e.g.~see \cite{Wilson2009a,Bohn2009a,Parker2009a,Billy2012a,Koch2008a,Lu2010a,Dutta2007a}). 
 For this untilted case the spectrum is isotropic, however this spectrum also applies along $k_y$ for other interaction parameter sets with $\alpha\ne0$, according to the mapping (\ref{e:map1})-(\ref{e:map2}). As an example we present a result for $\alpha=\frac{\pi}{3}$ in Fig.~\ref{f:chispec}(b). Along the $k_y$ axis the spectrum is the same as the untilted case, but along the $k_x$ axis the spectrum differs.\footnote{We note that $u_{k_y}^j(z)$, $v_{k_y}^j(z)$ can be taken real [e.g.~see Fig.~\ref{f:euvs}(b),(c)]. However, for $\alpha\ne 0$ and $k_x\ne0$, $u_{\bk_\rho}^j$, $v_{\bk_\rho}^j$ are generally complex.}

\begin{figure}[ht!] 
   \centering
  \includegraphics[width=3.4in]{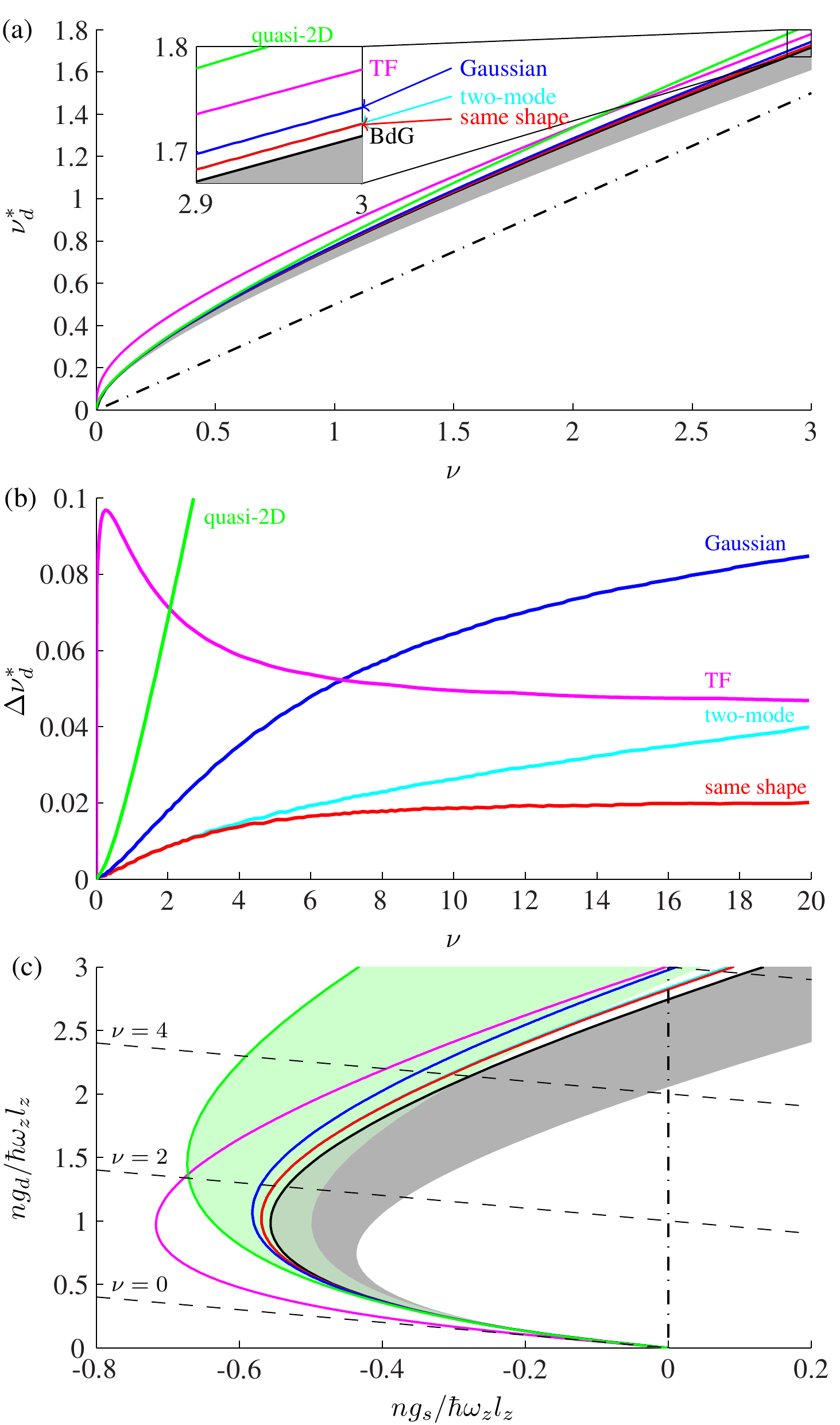}
  \caption{(a) Universal stability phase diagram for all $\alpha$, where $\nu_d^*$ is the critical value of $\nu_d$ where a quasiparticle mode becomes dynamically unstable. Boundary obtained numerically from full solution of the BdG equations (black, also the shaded roton region) and the same shape, variational, TF and quasi-2D approximations. The pure dipolar  (i.e.~$g_s=0$)  interaction line for $\alpha=0$ is shown for reference (dash-dotted), also see subplot (c).  (b) Excess of approximate stability result over the full solution of the GPE and quasi-particle equations. 
  (c) The stability diagram of subplot (a) mapped to $\alpha=0$ and plotted as a function of $ng_s/\hbar\omega_zl_z$, also showing the quasi-2D roton region (shaded green) and lines of constant $\nu$ (dashed) with $\nu=0$ being the phonon instability boundary. In these coordinates the pure dipolar line   (dash-dotted) is vertical. \label{f:stab2D} }
\end{figure}

As noted earlier, the $k_y$-spectrum is only a function of $\nu$ and $\nu_d$. Thus, using these as coordinates we can map out the stability and roton phase diagram for the system, with results shown in Fig.~\ref{f:stab2D}(a). In the shaded region the excitation spectrum is stable, but has a roton at   $k_y\ne {0}$.   We identify the stability boundary as where a quasiparticle mode has zero energy. That is, to the left of the stability boundary the mode softens  further and its eigenvalue $E_{k_y}$ becomes imaginary, signalling that the condensate is dynamically unstable.  We  emphasise that this phase diagram is universal for any tilt angle and the raw data for the stability boundary is available (see footnote 1). As an example, in Fig.~\ref{f:stab2D}(c) we show the phase diagram mapped to the untilted case as a function of  $g_s$ and $g_d$.

\section{Approximate treatments}\label{s:ApproxTreatments}
\subsection{General approximate schemes}
In what follows we investigate different approximate schemes for obtaining the condensate and excitations. These all have in common the assumption that the axial profile of the excitations is identical to the condensate, i.e. 
 \begin{equation}
       u_{\bk_\rho}(z)= \mathrm{u}_{\bk_\rho}\chi_\nu(z),\quad v_{\bk_\rho}(z)=\mathrm{v}_{\bk_\rho}\chi_\nu(z),\label{e:genss}
 \end{equation}
 where $\mathrm{u}_{\bk_\rho}$ and $\mathrm{v}_{\bk_\rho}$ are constants and we have dropped the superscript $j$, restricting ourselves to the ground mode in $z$. Our full numerical solutions for the $k_y$-quasiparticles [see Figs.~\ref{f:euvs}(b) and (c), particularly the insets] reveal that this is typically a good approximation for the $u$-modes, but is a poorer approximation for the $v$-modes.\footnote{Similar considerations are relevant for $k_x\ne0$ quasiparticles, where (for $\alpha\!\ne\!0$) $u_{\bk_\rho}$ and $v_{\bk_\rho}$ are irreducibly complex (c.f.~$\chi_\nu$ can be taken real).} This deformation of the $v$-modes tends to get worse for large DDIs, notably when a roton forms (hence near instability), but we note that 
 where the $v$-modes differ significantly in shape they tend to be relatively small in absolute norm, and thus it is not clear how important neglecting this feature will be.  For this reason we pay attention to how well the stability boundary is predicted by the various theories we derive based upon this assumption.

 The various approximate schemes we consider differ in the choice of $\chi_\nu$. Making assumption (\ref{e:genss}) allows us to integrate out the $z$ dimension, and reduce the problem to an effective 2D uniform system. Key to this simplification is the introduction  of the part of the effective 2D $k$-space interaction that does not contribute to the GPE (\ref{e:GPE1d}), i.e.\footnote{$F(\bk_\rho)$ is well behaved at $\alpha=\pi/2$ since $\nu_d$ includes a factor of $\cos^2\alpha$.}
\begin{align}
    F(\bk_\rho) &\equiv \frac{n}{\hbar\omega_z\gamma} \int \frac{dk_z}{2\pi}\,\tilde{U}(\bk)\left|\nt_\nu(k_z)\right|^2 - \nu,\\
  &=3\nu_d\left(\frac{k_x^2}{k_\rho^2}\tan^2\alpha-1\right) G_\nu\left(\frac{k_\rho l_z}{\sqrt{2}}\right),\label{UINTGen2D}
  \end{align}
  where $\nt_\nu(k_z)\equiv \mathcal{F}\{|\chi_\nu(z)|^2\}$ and $\tilde{U}(\bk)$ was given in Eq.~(\ref{Eqn:DDIk}).
We have also introduced
 \begin{align}
     G_\nu(q) &= \frac{1}{\gamma} \int \frac{dk_z}{2\pi}  \frac{l_z|\nt_\nu(k_z)|^2}{1+k_z^2l_z^2/(2q^2)}.\label{e:Gnu}
\end{align}
so that $G_\nu(0)=0$, $G_\nu(q)\to 1$ as $q\to\infty$ from \eqref{e:gamma} and, since the integrand is always positive, $G_\nu(q)$ is monotonically increasing in $q$. The spectrum is then obtained from the BdG equations~\eqref{e:BdG}
\begin{align}
    E_{\bk_\rho}^0 &= \sqrt{\frac{\hbar^2k_\rho^2}{2m}\left\{\frac{\hbar^2k_\rho^2}{2m}  + 2\gamma\hbar\omega_z[\nu +   F(\bk_\rho)]\right\}  }, \label{e:unifspec}
\end{align}
and the constants $\mathrm{u}_{\bk_\rho}$, $\mathrm{v}_{\bk_\rho}$ are given by
\begin{align}
    \mathrm{u}_{\bk_\rho} &=  \sqrt{\frac{\hbar^2 k_\rho^2/2m +  \gamma\hbar\omega_z[\nu +   F(\bk_\rho)]}{ 2E_{\bk_\rho}} + \frac12},\\
    \mathrm{v}_{\bk_\rho} &=  \mathrm{sgn}[\nu +   F(\bk_\rho)]\sqrt{\frac{\hbar^2 k_\rho^2/2m +  \gamma\hbar\omega_z[\nu +   F(\bk_\rho)]}{ 2E_{\bk_\rho}} - \frac12}. 
\end{align}
For $k_x=0$ or $\alpha=0$ the spectrum simplifies to 
\begin{align}
    E_{k_\rho}^0 &= \sqrt{\frac{\hbar^2k_\rho^2}{2m}\left\{\frac{\hbar^2k_\rho^2}{2m}  + 2\gamma\hbar\omega_z\left[\nu - 3\nu_d G_\nu\left(\frac{k_\rho l_z}{\sqrt{2}}\right)\right]\right\}  },
\end{align}
so that the interaction is always repulsive if $\nu_d<\nu/3$ and $\mathrm{v}_{k_\rho}$ becomes negative for sufficiently large $k_\rho$ if $\nu_d>\nu/3$.
\subsection{Same shape approximation}\label{S:sameshape}
The most immediate application is to use $\chi_\nu$ obtained from the numerical solution of the GPE mode to describe the axial shape of the quasiparticles. We refer to this as the \textit{same shape approximation}. This approach is hence fully numerical, but does not require numerical diagonalization, and provides a useful comparison to the other approximate schemes we now introduce. For numerical calculations (i.e. when not using the analytical results from the approximate schemes below), it is important to use $G_\nu(q)$ subject to a cutoff interaction which is shown in Appendix~\ref{s:CutoffInt}. We emphasise that because the same shape approximation involves the full GPE solution, all condensate quantities  (e.g.~$\bar{\mu},\gamma,\bar{E}_z$) are exact, and we use the term same shape to refer to the approximate description of the excitations.

\subsection{Variational and quasi-2D approximations\label{S:VarGau}}
An additional \emph{variational approximation} is to take $\chi_\nu$ to be an ansatz depending on one or more variational parameters to be determined by minimising the energy:
\begin{align}
      \frac{E[\chi_\nu]}{\hbar\omega_z}&=\int\!dz\,\chi_\nu^*\left(\hz + \frac{\nu}{2} l_z|\chi_\nu|^2\right)\chi_\nu,\\
       &=\Ebz(\nu)  + \frac{\gamma\nu}{2}.
\end{align}
The variational \emph{Gaussian} approximation is of the form
\begin{equation}
  \chi_\sigma(z)=\frac{e^{-z^2/2\sigma^2l_z^2}}{\pi^{1/4}\sqrt{\sigma l_z}},
\end{equation}
with length scale $\sigma l_z$ where $\sigma$ is a variational parameter, giving $\Ebz(\nu)  =1/(4\sigma^2)+\sigma^2/4$ and $\gamma = 1/(\sqrt{2\pi}\sigma )$. Using $\chi_\sigma$, the $k$-space interaction \eqref{UINTGen2D} can be evaluated with $G_\nu(q) = G_0(q\sigma)$ where 
\begin{align}
G_0(q) &=  \sqrt{\pi} q e^{q^2}\mathrm{erfc}(q),
\end{align}
with $\mathrm{erfc}$ the complementary error function. In general the optimal value of $\sigma$ is most easily determined numerically, however for small $\nu$
\begin{align}
\mub &\approx \frac{1}{2} + \frac{\nu}{\sqrt{2\pi}}-\frac{\nu^2}{32\pi},\label{e:muVApprox}\\
\gamma&\approx\frac{1}{\sqrt{2\pi}}-\frac{\nu}{8\pi}.\label{e:gammaVApprox}
\end{align} 
The quasi-2D approximation is when interactions are sufficiently weak ($\nu\to0$) that $\chi$ can be approximated as the non-interacting harmonic oscillator ground state with $\sigma=1$  (e.g.~see \cite{Ticknor2011a}).

A more complex but more accurate \emph{two-mode} variational approximation is based on the ground and second excited harmonic oscillator states 
\begin{equation}
    \chi_{\sigma,a_2}(z)=\frac{\chi_\sigma(z)}{\sqrt{1+a_2^2}}\left[1+\frac{a_2}{2\sqrt{2}} H_2(z/\sigma l_z)\right], \label{e:a2}
  \end{equation}
  where $H_2(u)=4u^2-2$ is the second Hermite polynomial and $a_2$ is an additional variational parameter. 
  This variational ansatz was first introduced by Wilson \textit{et al.}~\cite{Wilson2011a} for application to multilayered condensates (i.e.~sets of planar systems as produced by a one-dimensional optical lattice). 
  Expressions for $\gamma$, $\Ebz(\nu)$ and $G_{a_2}(q)$ [where $G_\nu(q) = G_{a_2}(q \sigma)$] using the two-mode approximation are given in Appendix \ref{s:a2appendix}.

\subsection{Axial TF approximation}\label{S:ATF} 
In the opposite regime to the quasi-2D case, i.e.~for $\nu\gg1$, interactions  dominate the confinement energy in the $z$ direction. In this regime a TF approximation is applicable (c.f.~\cite{Santos2003a})
\begin{equation}
|\chi_\mathrm{TF}(z)|^2=\frac{1}{\nu l_z}\max\left(0,\mub -z^2/2l_z^2\right),
\end{equation}
where
\begin{align}
\mub &=\frac{1}{2}\left(\frac{3\nu}{2}\right)^{2/3},\\
\gamma&=\frac{1}{5}\left(\frac{18}{\nu}\right)^{1/3}.
\end{align}
This ansatz  gives
\begin{align}
G^\mathrm{TF}_\nu(q)&=1- \frac{5}{2p^2}+15\frac{(p+1)^2e^{-2p}+p^2-1}{4p^5},
\end{align}
where $p\equiv\sqrt{2}qZ/l_z = 2q\sqrt{\mub}$, with $Z=\sqrt{2\mub}l_z$ the TF  half-width.

\subsection{Comparison of approximate schemes}

\begin{figure}[ht!] 
   \centering
  \includegraphics[width=3.4in]{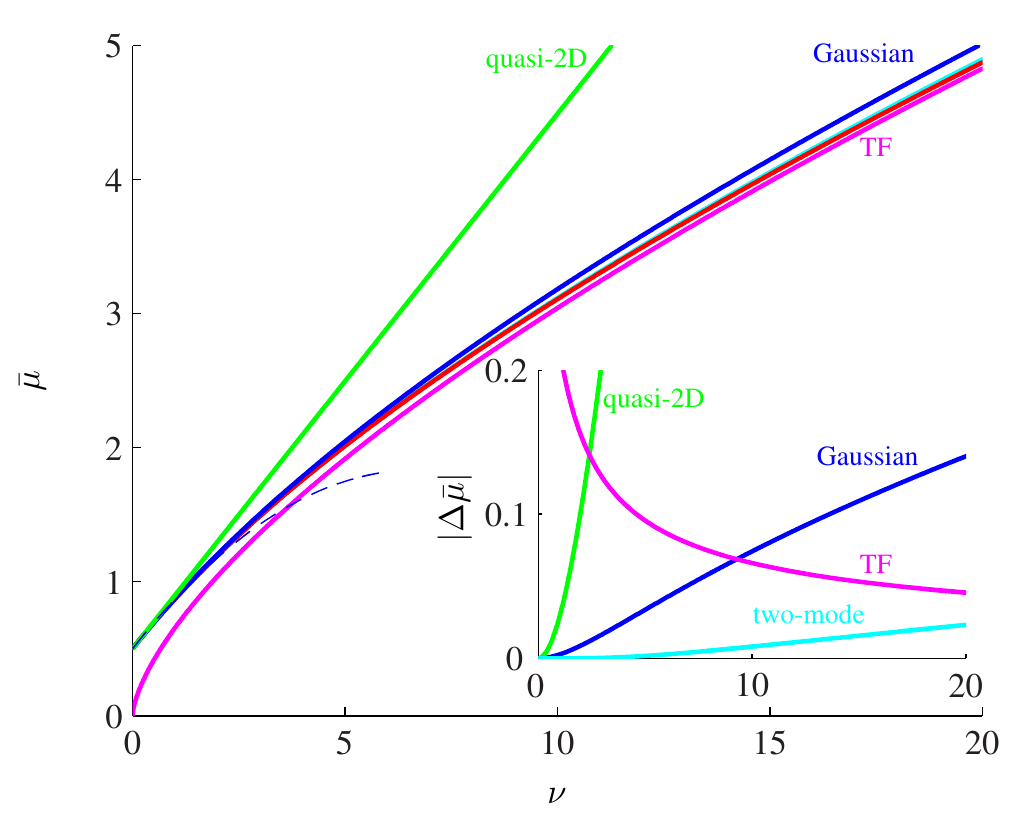}
  \caption{$\mub$ as a function of $\nu$ by solving the GPE \eqref{e:GPE1d} (red) and the variational, TF and quasi-2D $\mub = 1/2 + \nu/\sqrt{2\pi}$ approximations with the variational limit \eqref{e:muVApprox} (dashed).  Inset: absolute difference in $\mub$ between GPE and approximate result. The GPE result is bounded by the variational and TF results.\label{f:muGPE} } 
\end{figure}

\begin{figure}[ht!] 
   \centering
   \includegraphics[width=3.4in]{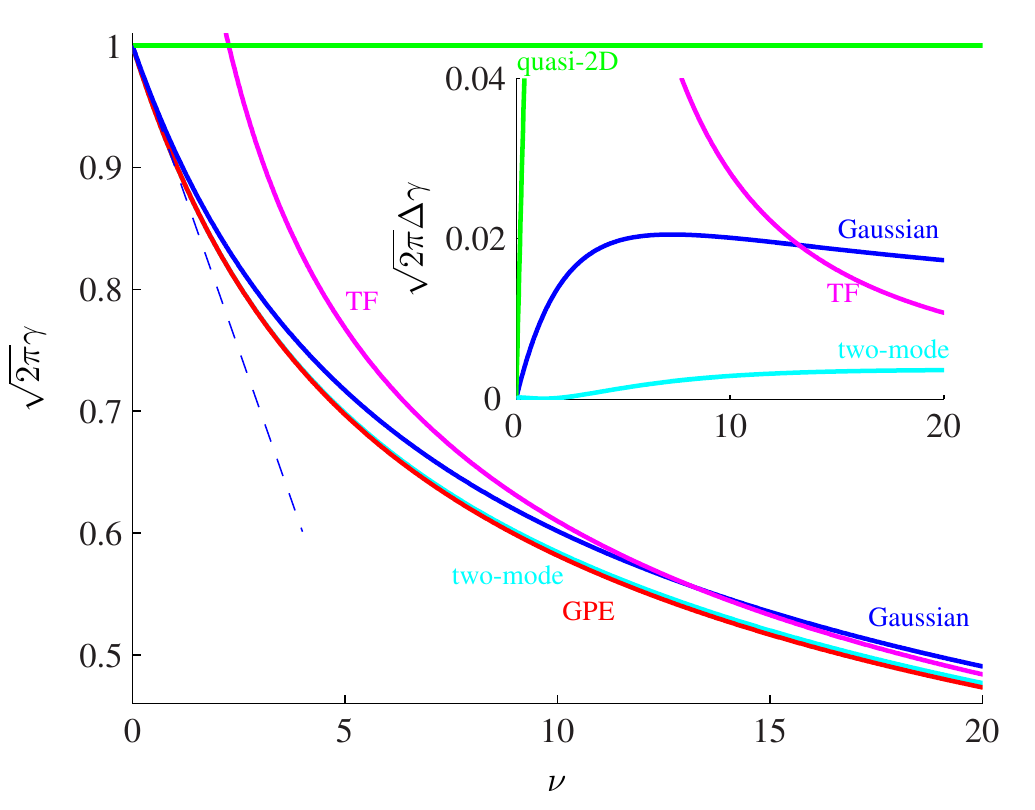}
   \caption{The dimensionless constant $\gamma$ as a function of $\nu$ by solving the GPE \eqref{e:GPE1d} and the variational, TF and quasi-2D approximations with the variational limit \eqref{e:gammaVApprox} (dashed). Inset: excess of approximate result over $\gamma$ from GPE.\label{f:gammaGPE} }
\end{figure}

\begin{figure}[ht!] 
   \centering
  \includegraphics[width=3.4in]{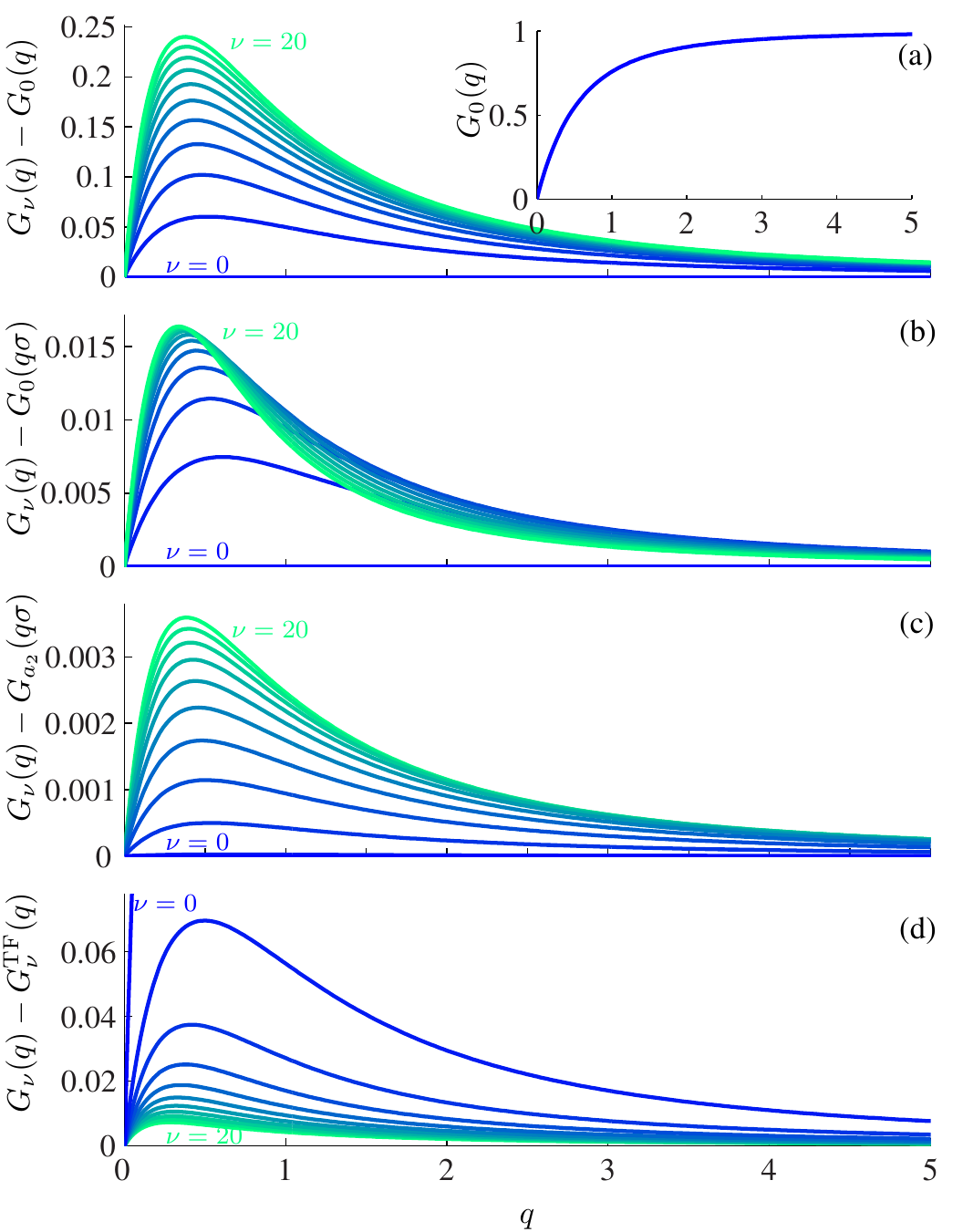}
  \caption{The various approximate $G_\nu$ functions are similar and are most easily distinguished by considering their differences from the same shape result as a function of $\nu$. (a)  $G_\nu(q)-G_{0}(q)$, (b) $G_\nu(q)-G_0( q\sigma)$, (c) $G_\nu(q)-G_{a_2}(q \sigma)$, (d) $G_\nu(q)-G^\mathrm{TF}_\nu(q)$ for $\nu$ from 0 (blue) to 20 (green) in steps of $2$. \label{f:GdiffGPE} 
  Inset: $G_0(q)$.}
\end{figure}

The approximate theories can be compared to each other by assessing the differences in their predictions for the functions $\bar{\mu}(\nu)$ [see Fig.~\ref{f:muGPE}], $\gamma(\nu)$  [see Fig.~\ref{f:gammaGPE}] and $G_\nu$ [see Fig.~\ref{f:GdiffGPE}], which are all presented over a broad range of $\nu$ values. These results make it clear that the quasi-2D approximation is only applicable for $\nu\ll1$. The two-mode variational results provides the best approximation for the range of interactions considered [e.g.~see insets to Figs.~\ref{f:muGPE} and \ref{f:gammaGPE}], although the TF approximation eventually becomes a better approximation at $\nu\sim25$. While the variational Gaussian is everywhere worse than the two-mode scheme, it is more accurate than the TF approximation until $\nu\sim10$.

The stability boundary is determined by the excitation spectrum, and the predictions of the various theories are compared against the full numerical result in Fig.~\ref{f:stab2D}.  Similar to our findings for $\bar\mu$ and $\gamma$, we see that the variational Gaussian is better than the axial TF description for interactions up to about $\nu\sim10$. However, the same shape approximation  always furnishes a better description than either the axial TF or variational Gaussian at all values of $\nu$ considered. We also consider the differences between $G_\nu$ (\ref{e:Gnu}), which determines the spectrum via (\ref{e:unifspec}), for the  approximate theories in Fig.~\ref{f:GdiffGPE}. Notably, Fig.~\ref{f:GdiffGPE}(b) shows that the difference between the same shape and variational Gaussian result increases with increasing $\nu$.
The two-mode variational result  [Fig.~\ref{f:GdiffGPE}(c)] has similar behavior, but the size of the difference is significantly smaller.
In Fig.~\ref{f:GdiffGPE}(d) the TF approximation is considered, and we observe that  the difference from the same shape results decreases with increasing $\nu$.

\begin{table}[htbp]
   \centering
   \begin{tabular}{@{} lcr @{}} 
      \toprule 
     Approach   & $\nu_d^*$\\
      \midrule
      BdG     & $2.743$  \\
      Same shape   &    $2.822$    \\
      Two-mode & $2.834$ \\
      Variational Gaussian      &  $2.974$   \\
      Axial TF &    $3.004$ \\
      Quasi-2D       &    $4.319$ \\
      \bottomrule
   \end{tabular}
   \caption{Critical value of interaction where a gas with $g_s=g_d\sin^2\alpha$, i.e. $\nu=2\nu_d$, becomes unstable. This includes the untilted purely dipolar gas.   \label{tab:critvals}}
\end{table}
 
In Fig.~\ref{f:chispec} we show the condensate solutions and excitation spectrum using the full and various approximate theories for an untilted purely dipolar case at instability (i.e.~at $\nu_d=2.743$ and $\nu=2\nu_d$). The spectrum shows that the full solution has a zero energy roton, whereas the various approximate theories have the roton minima at higher energies, except for the quasi-2D theory which does not have a roton for the interaction parameters considered. Increasing the dipole strength [i.e.~moving along the dash-dotted line in Fig.~\ref{f:stab2D}(a) or (c)] until each approximate theory predicts a zero energy roton we find their respective critical values $\nu_d^*$. These are tabulated in Table~\ref{tab:critvals} for comparison. The quasi-2D approximation is in worst agreement with the full theory and overestimates the stability value by a factor of about 1.6.

To put the results of this paper into context it is useful to consider parameters that are achievable in current experiments. Consider a condensate of $^{164}$Dy atoms \cite{Mingwu2011a}, which have a magnetic moment of $\mu=10\,\mu_B$ (where $\mu_B$ is the Bohr magneton). For the case of $N=1\times10^5$ condensed atoms confined in a pancake harmonic trap of frequencies $\{\omega_{\rho},\omega_z\}=2\pi\!\times\!\{10,500\}\,\mathrm{s}^{-1}$, and for untilted dipoles, we have $\nu_d=2.06$ and $\nu=5.69$ at trap center ($\brho=\mathbf{0}$).\footnote{To obtain this estimate we have taken the radial density profile to be described by a TF approximation  (c.f.~\cite{Parker2008a}), and we have assumed a background $s$-wave scattering length of 100 $a_0$ ($a_0$ is the Bohr radius).}

\section{Conclusions and Outlook}
In this paper we have studied the properties of a dipolar BEC flattened by confinement in the $z$ direction. We have pointed out the rigorous symmetries of the condensate and 
its excitation spectrum, which allow us to reduce the condensate solution to its dependence on a single effective interaction parameter $\nu$ that includes all aspects of the interactions. Similarly we find that the $k_y$-excitation spectrum, which is key to determining the emergence of rotons and stability, depends on two effective interaction parameters  $\{\nu,\nu_d\}$, allowing us (aided by exact numerical calculations) to provide a universal stability phase diagram in this parameter space.  

A second major focus of this paper has been to introduce and unify a number of analytic approximations for the condensate and excitations. The basis of these approaches consists of approximating the condensate profile in the confined direction by a specified profile. We introduced four analytic forms for this profile, some of which have been extensively used in the literature, particularly as a basis to simplify simulating the  Gross-Pitaevskii dynamics of the condensate.  These approximations are validated against our full numerical calculations (without approximation), and the same shape approximation. We validate the accuracy of these approximations by considering both condensate properties and the quasiparticle excitation spectrum. The best approximation is shown to be dependent on the strength of the interactions, and importantly that the widely used quasi-2D approximation is inaccurate except for very weak interactions where dipolar effects are small. 

The detailed understanding of the uniform planar case provided in this paper will underpin future work on tractable calculations of fully 3D trapped dipolar condensates without requiring cylindrical symmetry. This theory could be used to model the 3D trapped system with tight axial confinement and weaker in-plane trapping, by taking a wave function ansatz of the form 
\begin{equation}
\psi_0(\mathbf{r})=\sqrt{n(\bm{\rho})}\,\chi_{\nu(\bm{\rho})}(z),
\end{equation}
where $\nu$ depends on $\bm{\rho}$ via the areal density  $n(\brho)$ [see Eq.~\eqref{e:nu}]. With in-plane trapping, the radial Fourier transform of the density has a finite spread in $\mathbf{k}_{\bm{\rho}}$ [c.f.~the delta function in Eq.~\eqref{e:denFT} for uniform in-plane density]. 
The GPE then has two interaction terms, a contact term proportional to $\nu(\brho)$ locally [see Eq.~\eqref{e:LGPE}] and a long-range term depending on $n(\brho)$ everywhere. This long-range term has been discussed in the context of the quasi-2D approximation in Ref.~\cite{Cai2010a}, and can be important in weakly anisotropic traps (e.g.~$\omega_z\lesssim20\,\omega_{x,y}$, with $\omega_x$ and $\omega_y$ the radial trap frequencies) where it gives rise to density oscillating ground states \cite{Ronen2007a,Lu2010a}. For higher aspect ratio traps the radial density varies more slowly, and the long-range term becomes less significant. 

Stability of dipolar BECs has been of significant experimental interest \cite{Koch2008a}, and while dipole tilt is an easily accessed control parameter in experiments, accurate theory has been limited to situations of untilted dipoles (e.g.~see \cite{Ronen2007a,Lu2010a}) where the system has cylindrical symmetry. Our universal stability results fully quantify the role of tilt and present an opportunity for validation in current experiments. With radial trapping, the density changes with the interactions, notably the density will tend to decrease as the interactions increase because the condensate radially spreads out. The instability condition will occur when the maximum value of $\nu$  reaches the critical value predicted in the uniform case, although there will be corrections due to the effect of the radial trapping on the quasi-particle excitations.

Finally, we note that our work can be also extended down a number of other avenues, e.g., as the basis of descriptions for local collapse \cite{Linscott2014a} and condensate dynamics (e.g.~\cite{Edwards2012a}), the calculation of static \cite{Sykes2012a} and dynamic correlation functions, and finite temperature formalism for a trapped partially condensed dipolar condensate \cite{Ronen2007b,*Bisset2012,*Ticknor2012a}.

\section*{Acknowledgments:}
We thank R. N. Bisset for providing feedback on the manuscript. We acknowledge support by the Marsden Fund of the Royal Society of New Zealand (contract UOO1220).

\appendix
\section{Interaction potential for dipolar GPE}\label{s:intpotGPE}
We substitute $\psi_0(\br)=\sqrt{n}\chi_\nu(z)$ into Eq.~\eqref{e:fullGPE}. The Fourier transform of the condensate density is
\begin{equation}
\tilde{n}(\bk) =(2\pi)^2n\delta(\bk_\rho)\nt_\nu(k_z),\label{e:denFT}
\end{equation}
 so the interaction term gives
\begin{align}
    &    \int \frac{d\bk}{(2\pi)^3} e^{i \bk\cdot\bx}   \nt (\bk) \tilde{U}(\bk)= \hbar\omega_z\nu  l_z |\chi_\nu(z)|^2, \label{e:InterA}%\\
\end{align}
using \eqref{Eqn:DDIk} where $\nu$ is defined by \eqref{e:nu}. Since the condensate is uniform in the plane, only the $z$ kinetic energy contributes which leads to \eqref{e:GPE1d}. 
\section{Bogoliubov equations}\label{s:BogEqns}
The BdG equations are 
\begin{equation}
E_{\bk_\rho}^j   \begin{bmatrix} u_{\bk_\rho}^j(z)e^{i\bk_\rho\cdot\brho}
\\ v_{\bk_\rho}^j(z) e^{i\bk_\rho\cdot\brho}
 \end{bmatrix}    =   \begin{bmatrix}       \mathcal{L} & -\mathcal{M}\\
             \mathcal{M}^*  & -\mathcal{L}^* 
   \end{bmatrix}
     \begin{bmatrix} 
      u_{\bk_\rho}^j(z)e^{i\bk_\rho\cdot\brho}\\
     v_{\bk_\rho}^j(z)  e^{i\bk_\rho\cdot\brho}
 \end{bmatrix},
\end{equation}
where
\begin{align}
\mathcal{L}&\equiv \mathcal{L}_{\mathrm{GP}}-\frac{\hbar^2}{2m}\left(\frac{\partial^2}{\partial x^2}+\frac{\partial^2}{\partial y^2}\right)-\mu+\mathcal{M},\\
\mathcal{M} f(\br) &\equiv n\chi_\nu(z)\int d\br'\,  U(\br-\br')\chi_\nu(z')f(\br'),
\end{align}
so, using $\int d\br\,e^{-i(\bk'_\rho-\bk_\rho)\cdot\brho} f(z) = (2\pi)^2\delta(\bk_\rho-\bk'_\rho)\mathcal{F}\{f(z)\}$,
\begin{align}
\mathcal{L}[f^j_{\bk_\rho}(z)e^{i\bk_\rho\cdot\brho}] &= 
\left(\!\mathcal{L}_{\mathrm{GP}}+\frac{\hbar^2k_\rho^2}{2m}\!-\!\mu\!+\!\mathcal{M}\right)[f^j_{\bk_\rho}(z)e^{i\bk_\rho\cdot\brho}], \\
\mathcal{M}[f^j_{\bk_\rho}(z)e^{i\bk_\rho\cdot\brho}] 
&= n\chi_\nu(z)\mathcal{F}^{-1}\left\{ \Ut(\bk) \mathcal{F}\left\{ \chi_\nu(z) f^j_{\bk_\rho}(z)\right\}\right\}   e^{i \bk_\rho\cdot\brho },
\end{align}
so we may drop the $ e^{i \bk_\rho\cdot\brho }$ giving the simpler equations~\eqref{e:BdG}.
\section{Cutoff interaction} \label{s:CutoffInt}
Because we discretize with a grid of finite extent $Z$  in the $z$ direction, it is important to implement a cutoff interaction potential. This was first proposed for 3D trapped dipolar condensates in Appendix A of \cite{Ronen2006a}. 
Here we Fourier transform $U(\br)$ with a cutoff in position space (i.e.~set to zero for $|z|>Z$) giving 
\begin{align}
   \Ut^Z(\bk)\!  &=\!  \Ut(\bk) \!+\! 3g_d \cos^2\alpha e^{-Z k_\rho} 
   \left[ \left(1 \!  -\! \frac{k_x^2}{k_\rho^2}  \tan^2\alpha  \!\right)\!\frac{ k_\rho^2 \cos (Z k_z)\!-\!k_\rho k_z \sin (Z k_z)}{k_\rho^2 + k_z^2}  \! -\! 2\tan\alpha\frac{k_xk_\rho \sin(Z k_z) \!+\! k_x k_z\cos(Z k_z)}{k_\rho^2 + k_z^2}  \right]\!,
   \end{align}
which reduces to the result of \cite{Ronen2006a} for $\alpha=0$, or (with a factor of $\cos^2\alpha$) when $k_x=0$. We also find
\begin{align}
    G_\nu^Z(q) &= G_\nu(q) - e^{-\sqrt{2}Zq/l_z}\frac{l_z}{\gamma}\int \frac{dk_z}{2\pi}\frac{|\nt_\nu(k_z)|^2}{q^2+k_z^2l_z^2/2}  \left[q^2 \cos (Zk_z) - \frac{qk_zl_z}{\sqrt2} \sin (Zk_z) \right].  \label{e:GZcut}
\end{align}
\section{Two-mode variational approximation} \label{s:a2appendix}
 \begin{figure}[t!] 
   \centering 
   \includegraphics[width=3.5in]{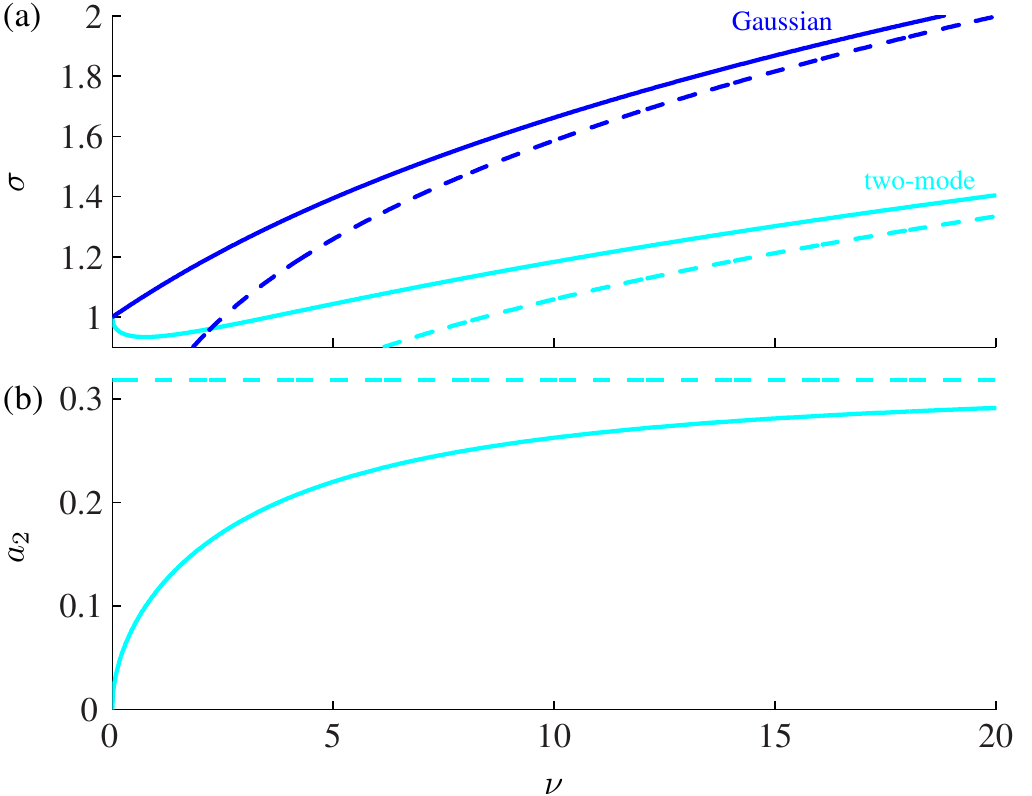}
   \caption{ Variational parameters (a) $\sigma$ for Gaussian (blue) with the large $\nu$ limit $\sigma \approx \nu^{1/3}/(2\pi)^{1/6}$ (blue dashed) and two-mode (cyan) with the large $\nu$ limit $\sigma \approx 0.4916\nu^{1/3}$ (cyan dashed) and (b) $a_2$ and the large $\nu$ limit $a_2\approx 0.3186$ (dashed).
   \label{f:a2} }    
\end{figure}
Using the two-mode variational approximation \eqref{e:a2}, the expressions for $\gamma$, $\Ebz(\nu)$ and $G_{a_2}(q)$ are
\begin{align}
     \gamma &= \frac{1 - \sqrt{2} a_2 +\frac94 a_2^2 +\frac{1}{4\sqrt{2}} a_2^3 + \frac{41}{64} a_2^4}{\sqrt{2\pi} \sigma(1+a_2^2)^2},\\
    \Ebz(\nu) &= \frac{1}{1+a_2^2}\left[ \frac{1}{4\sigma^2}(1-2\sqrt{2} a_2+5a_2^2) + \frac{\sigma^2}{4}(1+2\sqrt{2} a_2+5a_2^2)\right],\\
    G_{a_2}(q)  &=  \frac{\frac{a_2^4}{4} \left(q^4+4 q^2+2\right)^2+\sqrt{2} a_2^3 q^2
   \left(q^4+4 q^2+2\right)+ a_2^2 \left(3 q^4+4 q^2+2\right)+2 \sqrt{2} a_2 q^2+1 }
   {1 - \sqrt{2} a_2 +\frac94 a_2^2 +\frac{1}{4\sqrt{2}} a_2^3 + \frac{41}{64} a_2^4} G_0(q)\notag\\
   &-
   \frac{ \frac{a_2^3}{32} \left(8 q^6+60 q^4+134 q^2+81\right)
   +\frac{a_2^2}{2 \sqrt{2}} \left(4 q^4+14 q^2+3\right)+\frac{a_2}{2} \left(6 q^2+5\right)+ 2\sqrt{2}}
   {1 - \sqrt{2} a_2 +\frac94 a_2^2 +\frac{1}{4\sqrt{2}} a_2^3 + \frac{41}{64} a_2^4} a_2 q^2.
   \end{align}
Using the notation $F_\mathrm{eff}(q)$ from \cite{Wilson2011a} we can write $G_{a_2}(q) =  \frac{2}{3} \left[1-\frac{F_\mathrm{eff}(q)}{F_\mathrm{eff}(0)}\right]$. The variational parameters $\sigma$ and $a_2$ must be determined numerically with results shown in Fig.~\ref{f:a2}. The two-mode variational approximation develops a density oscillation in the $z$ direction when $a_2 > \sqrt{2}/5 \approx 0.28$ (Fig.~\ref{f:a2} shows this $a_2$ arising for $\nu \gtrsim 16$). This is a symptom of the beginning of the breakdown of the two-mode approximation. The solution to the GPE, Eq.~\eqref{e:GPE1d}, never develops a density oscillation. For all $\nu$ where the two-mode approximation is better than the TF approximation we have $\max_z[\chi_{\sigma,a_2}(z)]/\chi_{\sigma,a_2}(0)-1<1\%$.

\bibliographystyle{apsrev4-1}
%\bibliography{../dipolarpbb}

\begin{thebibliography}{58}%
\makeatletter
\providecommand \@ifxundefined [1]{%
 \@ifx{#1\undefined}
}%
\providecommand \@ifnum [1]{%
 \ifnum #1\expandafter \@firstoftwo
 \else \expandafter \@secondoftwo
 \fi
}%
\providecommand \@ifx [1]{%
 \ifx #1\expandafter \@firstoftwo
 \else \expandafter \@secondoftwo
 \fi
}%
\providecommand \natexlab [1]{#1}%
\providecommand \enquote  [1]{``#1''}%
\providecommand \bibnamefont  [1]{#1}%
\providecommand \bibfnamefont [1]{#1}%
\providecommand \citenamefont [1]{#1}%
\providecommand \href@noop [0]{\@secondoftwo}%
\providecommand \href [0]{\begingroup \@sanitize@url \@href}%
\providecommand \@href[1]{\@@startlink{#1}\@@href}%
\providecommand \@@href[1]{\endgroup#1\@@endlink}%
\providecommand \@sanitize@url [0]{\catcode `\\12\catcode `\$12\catcode
  `\&12\catcode `\#12\catcode `\^12\catcode `\_12\catcode `\%12\relax}%
\providecommand \@@startlink[1]{}%
\providecommand \@@endlink[0]{}%
\providecommand \url  [0]{\begingroup\@sanitize@url \@url }%
\providecommand \@url [1]{\endgroup\@href {#1}{\urlprefix }}%
\providecommand \urlprefix  [0]{URL }%
\providecommand \Eprint [0]{\href }%
\providecommand \doibase [0]{http://dx.doi.org/}%
\providecommand \selectlanguage [0]{\@gobble}%
\providecommand \bibinfo  [0]{\@secondoftwo}%
\providecommand \bibfield  [0]{\@secondoftwo}%
\providecommand \translation [1]{[#1]}%
\providecommand \BibitemOpen [0]{}%
\providecommand \bibitemStop [0]{}%
\providecommand \bibitemNoStop [0]{.\EOS\space}%
\providecommand \EOS [0]{\spacefactor3000\relax}%
\providecommand \BibitemShut  [1]{\csname bibitem#1\endcsname}%
\let\auto@bib@innerbib\@empty
%</preamble>
\bibitem [{\citenamefont {Griesmaier}\ \emph {et~al.}(2005)\citenamefont
  {Griesmaier}, \citenamefont {Werner}, \citenamefont {Hensler}, \citenamefont
  {Stuhler},\ and\ \citenamefont {Pfau}}]{Griesmaier2005a}%
  \BibitemOpen
  \bibfield  {author} {\bibinfo {author} {\bibfnamefont {A.}~\bibnamefont
  {Griesmaier}}, \bibinfo {author} {\bibfnamefont {J.}~\bibnamefont {Werner}},
  \bibinfo {author} {\bibfnamefont {S.}~\bibnamefont {Hensler}}, \bibinfo
  {author} {\bibfnamefont {J.}~\bibnamefont {Stuhler}}, \ and\ \bibinfo
  {author} {\bibfnamefont {T.}~\bibnamefont {Pfau}},\ }\href {\doibase %
  10.1103/PhysRevLett.94.160401} {\bibfield  {journal} {\bibinfo  {journal}
  {Phys. Rev. Lett.}\ }\textbf {\bibinfo {volume} {94}},\ \bibinfo {pages}
  {160401} (\bibinfo {year} {2005})}\BibitemShut {NoStop}%
\bibitem [{\citenamefont {Beaufils}\ \emph {et~al.}(2008)\citenamefont
  {Beaufils}, \citenamefont {Chicireanu}, \citenamefont {Zanon}, \citenamefont
  {Laburthe-Tolra}, \citenamefont {Mar\'echal}, \citenamefont {Vernac},
  \citenamefont {Keller},\ and\ \citenamefont {Gorceix}}]{Beaufils2008a}%
  \BibitemOpen
  \bibfield  {author} {\bibinfo {author} {\bibfnamefont {Q.}~\bibnamefont
  {Beaufils}}, \bibinfo {author} {\bibfnamefont {R.}~\bibnamefont
  {Chicireanu}}, \bibinfo {author} {\bibfnamefont {T.}~\bibnamefont {Zanon}},
  \bibinfo {author} {\bibfnamefont {B.}~\bibnamefont {Laburthe-Tolra}},
  \bibinfo {author} {\bibfnamefont {E.}~\bibnamefont {Mar\'echal}}, \bibinfo
  {author} {\bibfnamefont {L.}~\bibnamefont {Vernac}}, \bibinfo {author}
  {\bibfnamefont {J.-C.}\ \bibnamefont {Keller}}, \ and\ \bibinfo {author}
  {\bibfnamefont {O.}~\bibnamefont {Gorceix}},\ }\href {\doibase %
  10.1103/PhysRevA.77.061601} {\bibfield  {journal} {\bibinfo  {journal} {Phys.
  Rev. A}\ }\textbf {\bibinfo {volume} {77}},\ \bibinfo {pages} {061601}
  (\bibinfo {year} {2008})}\BibitemShut {NoStop}%
\bibitem [{\citenamefont {Lu}\ \emph {et~al.}(2011)\citenamefont {Lu},
  \citenamefont {Burdick}, \citenamefont {Youn},\ and\ \citenamefont
  {Lev}}]{Mingwu2011a}%
  \BibitemOpen
  \bibfield  {author} {\bibinfo {author} {\bibfnamefont {M.}~\bibnamefont
  {Lu}}, \bibinfo {author} {\bibfnamefont {N.~Q.}\ \bibnamefont {Burdick}},
  \bibinfo {author} {\bibfnamefont {S.~H.}\ \bibnamefont {Youn}}, \ and\
  \bibinfo {author} {\bibfnamefont {B.~L.}\ \bibnamefont {Lev}},\ }\href
  {\doibase 10.1103/PhysRevLett.107.190401} {\bibfield  {journal} {\bibinfo
  {journal} {Phys. Rev. Lett.}\ }\textbf {\bibinfo {volume} {107}},\ \bibinfo
  {pages} {190401} (\bibinfo {year} {2011})}\BibitemShut {NoStop}%
\bibitem [{\citenamefont {Aikawa}\ \emph {et~al.}(2012)\citenamefont {Aikawa},
  \citenamefont {Frisch}, \citenamefont {Mark}, \citenamefont {Baier},
  \citenamefont {Rietzler}, \citenamefont {Grimm},\ and\ \citenamefont
  {Ferlaino}}]{Aikawa2012a}%
  \BibitemOpen
  \bibfield  {author} {\bibinfo {author} {\bibfnamefont {K.}~\bibnamefont
  {Aikawa}}, \bibinfo {author} {\bibfnamefont {A.}~\bibnamefont {Frisch}},
  \bibinfo {author} {\bibfnamefont {M.}~\bibnamefont {Mark}}, \bibinfo {author}
  {\bibfnamefont {S.}~\bibnamefont {Baier}}, \bibinfo {author} {\bibfnamefont
  {A.}~\bibnamefont {Rietzler}}, \bibinfo {author} {\bibfnamefont
  {R.}~\bibnamefont {Grimm}}, \ and\ \bibinfo {author} {\bibfnamefont
  {F.}~\bibnamefont {Ferlaino}},\ }\href {\doibase %
  10.1103/PhysRevLett.108.210401} {\bibfield  {journal} {\bibinfo  {journal}
  {Phys. Rev. Lett.}\ }\textbf {\bibinfo {volume} {108}},\ \bibinfo {pages}
  {210401} (\bibinfo {year} {2012})}\BibitemShut {NoStop}%
\bibitem [{\citenamefont {Baranov}(2008)}]{Baranov2008}%
  \BibitemOpen
  \bibfield  {author} {\bibinfo {author} {\bibfnamefont {M.}~\bibnamefont
  {Baranov}},\ }\href {\doibase 10.1016/j.physrep.2008.04.007} {\bibfield
  {journal} {\bibinfo  {journal} {Phys. Rep.}\ }\textbf {\bibinfo {volume}
  {464}},\ \bibinfo {pages} {71 } (\bibinfo {year} {2008})}\BibitemShut
  {NoStop}%
\bibitem [{\citenamefont {Lahaye}\ \emph {et~al.}(2009)\citenamefont {Lahaye},
  \citenamefont {Menotti}, \citenamefont {Santos}, \citenamefont {Lewenstein},\
  and\ \citenamefont {Pfau}}]{Lahaye_RepProgPhys_2009}%
  \BibitemOpen
  \bibfield  {author} {\bibinfo {author} {\bibfnamefont {T.}~\bibnamefont
  {Lahaye}}, \bibinfo {author} {\bibfnamefont {C.}~\bibnamefont {Menotti}},
  \bibinfo {author} {\bibfnamefont {L.}~\bibnamefont {Santos}}, \bibinfo
  {author} {\bibfnamefont {M.}~\bibnamefont {Lewenstein}}, \ and\ \bibinfo
  {author} {\bibfnamefont {T.}~\bibnamefont {Pfau}},\ }\href
  {http://stacks.iop.org/0034-4885/72/126401} {\bibfield  {journal} {\bibinfo
  {journal} {Rep. Prog. Phys.}\ }\textbf {\bibinfo {volume} {72}},\ \bibinfo
  {pages} {126401} (\bibinfo {year} {2009})}\BibitemShut {NoStop}%
\bibitem [{\citenamefont {Koch}\ \emph {et~al.}(2008)\citenamefont {Koch},
  \citenamefont {Lahaye}, \citenamefont {Metz}, \citenamefont {Frohlich},
  \citenamefont {Griesmaier},\ and\ \citenamefont {Pfau}}]{Koch2008a}%
  \BibitemOpen
  \bibfield  {author} {\bibinfo {author} {\bibfnamefont {T.}~\bibnamefont
  {Koch}}, \bibinfo {author} {\bibfnamefont {T.}~\bibnamefont {Lahaye}},
  \bibinfo {author} {\bibfnamefont {J.}~\bibnamefont {Metz}}, \bibinfo {author}
  {\bibfnamefont {B.}~\bibnamefont {Frohlich}}, \bibinfo {author}
  {\bibfnamefont {A.}~\bibnamefont {Griesmaier}}, \ and\ \bibinfo {author}
  {\bibfnamefont {T.}~\bibnamefont {Pfau}},\ }\href {\doibase 10.1038/nphys887}
  {\bibfield  {journal} {\bibinfo  {journal} {Nat. Phys.}\ }\textbf {\bibinfo
  {volume} {4}},\ \bibinfo {pages} {218} (\bibinfo {year} {2008})}\BibitemShut
  {NoStop}%
\bibitem [{\citenamefont {M\"uller}\ \emph {et~al.}(2011)\citenamefont
  {M\"uller}, \citenamefont {Billy}, \citenamefont {Henn}, \citenamefont
  {Kadau}, \citenamefont {Griesmaier}, \citenamefont {Jona-Lasinio},
  \citenamefont {Santos},\ and\ \citenamefont {Pfau}}]{Muller2011a}%
  \BibitemOpen
  \bibfield  {author} {\bibinfo {author} {\bibfnamefont {S.}~\bibnamefont
  {M\"uller}}, \bibinfo {author} {\bibfnamefont {J.}~\bibnamefont {Billy}},
  \bibinfo {author} {\bibfnamefont {E.~A.~L.}\ \bibnamefont {Henn}}, \bibinfo
  {author} {\bibfnamefont {H.}~\bibnamefont {Kadau}}, \bibinfo {author}
  {\bibfnamefont {A.}~\bibnamefont {Griesmaier}}, \bibinfo {author}
  {\bibfnamefont {M.}~\bibnamefont {Jona-Lasinio}}, \bibinfo {author}
  {\bibfnamefont {L.}~\bibnamefont {Santos}}, \ and\ \bibinfo {author}
  {\bibfnamefont {T.}~\bibnamefont {Pfau}},\ }\href {\doibase %
  10.1103/PhysRevA.84.053601} {\bibfield  {journal} {\bibinfo  {journal} {Phys.
  Rev. A}\ }\textbf {\bibinfo {volume} {84}},\ \bibinfo {pages} {053601}
  (\bibinfo {year} {2011})}\BibitemShut {NoStop}%
\bibitem [{\citenamefont {Ronen}\ \emph {et~al.}(2007)\citenamefont {Ronen},
  \citenamefont {Bortolotti},\ and\ \citenamefont {Bohn}}]{Ronen2007a}%
  \BibitemOpen
  \bibfield  {author} {\bibinfo {author} {\bibfnamefont {S.}~\bibnamefont
  {Ronen}}, \bibinfo {author} {\bibfnamefont {D.~C.~E.}\ \bibnamefont
  {Bortolotti}}, \ and\ \bibinfo {author} {\bibfnamefont {J.~L.}\ \bibnamefont
  {Bohn}},\ }\href {\doibase 10.1103/PhysRevLett.98.030406} {\bibfield
  {journal} {\bibinfo  {journal} {Phys. Rev. Lett.}\ }\textbf {\bibinfo
  {volume} {98}},\ \bibinfo {eid} {030406} (\bibinfo {year}
  {2007})}\BibitemShut {NoStop}%
\bibitem [{\citenamefont {Lu}\ \emph {et~al.}(2010)\citenamefont {Lu},
  \citenamefont {Lu}, \citenamefont {Zhang}, \citenamefont {Qiu}, \citenamefont
  {Pu},\ and\ \citenamefont {Yi}}]{Lu2010a}%
  \BibitemOpen
  \bibfield  {author} {\bibinfo {author} {\bibfnamefont {H.-Y.}\ \bibnamefont
  {Lu}}, \bibinfo {author} {\bibfnamefont {H.}~\bibnamefont {Lu}}, \bibinfo
  {author} {\bibfnamefont {J.-N.}\ \bibnamefont {Zhang}}, \bibinfo {author}
  {\bibfnamefont {R.-Z.}\ \bibnamefont {Qiu}}, \bibinfo {author} {\bibfnamefont
  {H.}~\bibnamefont {Pu}}, \ and\ \bibinfo {author} {\bibfnamefont
  {S.}~\bibnamefont {Yi}},\ }\href {\doibase 10.1103/PhysRevA.82.023622}
  {\bibfield  {journal} {\bibinfo  {journal} {Phys. Rev. A}\ }\textbf {\bibinfo
  {volume} {82}},\ \bibinfo {pages} {023622} (\bibinfo {year}
  {2010})}\BibitemShut {NoStop}%
\bibitem [{\citenamefont {Asad-uz Zaman}\ and\ \citenamefont
  {Blume}(2010)}]{Asad-uz-Zaman2010a}%
  \BibitemOpen
  \bibfield  {author} {\bibinfo {author} {\bibfnamefont {M.}~\bibnamefont
  {Asad-uz Zaman}}\ and\ \bibinfo {author} {\bibfnamefont {D.}~\bibnamefont
  {Blume}},\ }\href {http://stacks.iop.org/1367-2630/12/i=6/a=065022}
  {\bibfield  {journal} {\bibinfo  {journal} {New J. Phys.}\ }\textbf {\bibinfo
  {volume} {12}},\ \bibinfo {pages} {065022} (\bibinfo {year}
  {2010})}\BibitemShut {NoStop}%
\bibitem [{\citenamefont {Martin}\ and\ \citenamefont
  {Blakie}(2012)}]{Martin2012a}%
  \BibitemOpen
  \bibfield  {author} {\bibinfo {author} {\bibfnamefont {A.~D.}\ \bibnamefont
  {Martin}}\ and\ \bibinfo {author} {\bibfnamefont {P.~B.}\ \bibnamefont
  {Blakie}},\ }\href {\doibase 10.1103/PhysRevA.86.053623} {\bibfield
  {journal} {\bibinfo  {journal} {Phys. Rev. A}\ }\textbf {\bibinfo {volume}
  {86}},\ \bibinfo {pages} {053623} (\bibinfo {year} {2012})}\BibitemShut
  {NoStop}%
\bibitem [{\citenamefont {Bisset}\ and\ \citenamefont
  {Blakie}(2013)}]{Bisset2013a}%
  \BibitemOpen
  \bibfield  {author} {\bibinfo {author} {\bibfnamefont {R.~N.}\ \bibnamefont
  {Bisset}}\ and\ \bibinfo {author} {\bibfnamefont {P.~B.}\ \bibnamefont
  {Blakie}},\ }\href {\doibase 10.1103/PhysRevLett.110.265302} {\bibfield
  {journal} {\bibinfo  {journal} {Phys. Rev. Lett.}\ }\textbf {\bibinfo
  {volume} {110}},\ \bibinfo {pages} {265302} (\bibinfo {year}
  {2013})}\BibitemShut {NoStop}%
\bibitem [{\citenamefont {Boudjem\^aa}\ and\ \citenamefont
  {Shlyapnikov}(2013)}]{Boudjemaa2013a}%
  \BibitemOpen
  \bibfield  {author} {\bibinfo {author} {\bibfnamefont {A.}~\bibnamefont
  {Boudjem\^aa}}\ and\ \bibinfo {author} {\bibfnamefont {G.~V.}\ \bibnamefont
  {Shlyapnikov}},\ }\href {\doibase 10.1103/PhysRevA.87.025601} {\bibfield
  {journal} {\bibinfo  {journal} {Phys. Rev. A}\ }\textbf {\bibinfo {volume}
  {87}},\ \bibinfo {pages} {025601} (\bibinfo {year} {2013})}\BibitemShut
  {NoStop}%
\bibitem [{\citenamefont {Blakie}\ \emph {et~al.}(2013)\citenamefont {Blakie},
  \citenamefont {Baillie},\ and\ \citenamefont {Bisset}}]{Blakie2013a}%
  \BibitemOpen
  \bibfield  {author} {\bibinfo {author} {\bibfnamefont {P.~B.}\ \bibnamefont
  {Blakie}}, \bibinfo {author} {\bibfnamefont {D.}~\bibnamefont {Baillie}}, \
  and\ \bibinfo {author} {\bibfnamefont {R.~N.}\ \bibnamefont {Bisset}},\
  }\href {\doibase 10.1103/PhysRevA.88.013638} {\bibfield  {journal} {\bibinfo
  {journal} {Phys. Rev. A}\ }\textbf {\bibinfo {volume} {88}},\ \bibinfo
  {pages} {013638} (\bibinfo {year} {2013})}\BibitemShut {NoStop}%
\bibitem [{\citenamefont {Santos}\ \emph {et~al.}(2003)\citenamefont {Santos},
  \citenamefont {Shlyapnikov},\ and\ \citenamefont {Lewenstein}}]{Santos2003a}%
  \BibitemOpen
  \bibfield  {author} {\bibinfo {author} {\bibfnamefont {L.}~\bibnamefont
  {Santos}}, \bibinfo {author} {\bibfnamefont {G.~V.}\ \bibnamefont
  {Shlyapnikov}}, \ and\ \bibinfo {author} {\bibfnamefont {M.}~\bibnamefont
  {Lewenstein}},\ }\href {\doibase 10.1103/PhysRevLett.90.250403} {\bibfield
  {journal} {\bibinfo  {journal} {Phys. Rev. Lett.}\ }\textbf {\bibinfo
  {volume} {90}},\ \bibinfo {pages} {250403} (\bibinfo {year}
  {2003})}\BibitemShut {NoStop}%
\bibitem [{\citenamefont {Nath}\ and\ \citenamefont
  {Santos}(2010)}]{Nath2010a}%
  \BibitemOpen
  \bibfield  {author} {\bibinfo {author} {\bibfnamefont {R.}~\bibnamefont
  {Nath}}\ and\ \bibinfo {author} {\bibfnamefont {L.}~\bibnamefont {Santos}},\
  }\href {\doibase 10.1103/PhysRevA.81.033626} {\bibfield  {journal} {\bibinfo
  {journal} {Phys. Rev. A}\ }\textbf {\bibinfo {volume} {81}},\ \bibinfo
  {pages} {033626} (\bibinfo {year} {2010})}\BibitemShut {NoStop}%
\bibitem [{\citenamefont {Hufnagl}\ \emph {et~al.}(2011)\citenamefont
  {Hufnagl}, \citenamefont {Kaltseis}, \citenamefont {Apaja},\ and\
  \citenamefont {Zillich}}]{Hufnagl2011a}%
  \BibitemOpen
  \bibfield  {author} {\bibinfo {author} {\bibfnamefont {D.}~\bibnamefont
  {Hufnagl}}, \bibinfo {author} {\bibfnamefont {R.}~\bibnamefont {Kaltseis}},
  \bibinfo {author} {\bibfnamefont {V.}~\bibnamefont {Apaja}}, \ and\ \bibinfo
  {author} {\bibfnamefont {R.~E.}\ \bibnamefont {Zillich}},\ }\href {\doibase %
  10.1103/PhysRevLett.107.065303} {\bibfield  {journal} {\bibinfo  {journal}
  {Phys. Rev. Lett.}\ }\textbf {\bibinfo {volume} {107}},\ \bibinfo {pages}
  {065303} (\bibinfo {year} {2011})}\BibitemShut {NoStop}%
\bibitem [{\citenamefont {Blakie}\ \emph {et~al.}(2012)\citenamefont {Blakie},
  \citenamefont {Baillie},\ and\ \citenamefont {Bisset}}]{Blakie2012a}%
  \BibitemOpen
  \bibfield  {author} {\bibinfo {author} {\bibfnamefont {P.~B.}\ \bibnamefont
  {Blakie}}, \bibinfo {author} {\bibfnamefont {D.}~\bibnamefont {Baillie}}, \
  and\ \bibinfo {author} {\bibfnamefont {R.~N.}\ \bibnamefont {Bisset}},\
  }\href {\doibase 10.1103/PhysRevA.86.021604} {\bibfield  {journal} {\bibinfo
  {journal} {Phys. Rev. A}\ }\textbf {\bibinfo {volume} {86}},\ \bibinfo
  {pages} {021604} (\bibinfo {year} {2012})}\BibitemShut {NoStop}%
\bibitem [{\citenamefont {Jona-Lasinio}\ \emph {et~al.}(2013)\citenamefont
  {Jona-Lasinio}, \citenamefont {\L{}akomy},\ and\ \citenamefont
  {Santos}}]{JonaLasinio2013}%
  \BibitemOpen
  \bibfield  {author} {\bibinfo {author} {\bibfnamefont {M.}~\bibnamefont
  {Jona-Lasinio}}, \bibinfo {author} {\bibfnamefont {K.}~\bibnamefont
  {\L{}akomy}}, \ and\ \bibinfo {author} {\bibfnamefont {L.}~\bibnamefont
  {Santos}},\ }\href {\doibase 10.1103/PhysRevA.88.013619} {\bibfield
  {journal} {\bibinfo  {journal} {Phys. Rev. A}\ }\textbf {\bibinfo {volume}
  {88}},\ \bibinfo {pages} {013619} (\bibinfo {year} {2013})}\BibitemShut
  {NoStop}%
\bibitem [{\citenamefont {Bisset}\ \emph {et~al.}(2013)\citenamefont {Bisset},
  \citenamefont {Baillie},\ and\ \citenamefont {Blakie}}]{Bisset2013b}%
  \BibitemOpen
  \bibfield  {author} {\bibinfo {author} {\bibfnamefont {R.~N.}\ \bibnamefont
  {Bisset}}, \bibinfo {author} {\bibfnamefont {D.}~\bibnamefont {Baillie}}, \
  and\ \bibinfo {author} {\bibfnamefont {P.~B.}\ \bibnamefont {Blakie}},\
  }\href {\doibase 10.1103/PhysRevA.88.043606} {\bibfield  {journal} {\bibinfo
  {journal} {Phys. Rev. A}\ }\textbf {\bibinfo {volume} {88}},\ \bibinfo
  {pages} {043606} (\bibinfo {year} {2013})}\BibitemShut {NoStop}%
\bibitem [{\citenamefont {Wilson}\ \emph {et~al.}(2010)\citenamefont {Wilson},
  \citenamefont {Ronen},\ and\ \citenamefont {Bohn}}]{Wilson2010a}%
  \BibitemOpen
  \bibfield  {author} {\bibinfo {author} {\bibfnamefont {R.~M.}\ \bibnamefont
  {Wilson}}, \bibinfo {author} {\bibfnamefont {S.}~\bibnamefont {Ronen}}, \
  and\ \bibinfo {author} {\bibfnamefont {J.~L.}\ \bibnamefont {Bohn}},\ }\href
  {\doibase 10.1103/PhysRevLett.104.094501} {\bibfield  {journal} {\bibinfo
  {journal} {Phys. Rev. Lett.}\ }\textbf {\bibinfo {volume} {104}},\ \bibinfo
  {pages} {094501} (\bibinfo {year} {2010})}\BibitemShut {NoStop}%
\bibitem [{\citenamefont {Giovanazzi}\ \emph {et~al.}(2002)\citenamefont
  {Giovanazzi}, \citenamefont {G\"orlitz},\ and\ \citenamefont
  {Pfau}}]{Giovanazzi2002a}%
  \BibitemOpen
  \bibfield  {author} {\bibinfo {author} {\bibfnamefont {S.}~\bibnamefont
  {Giovanazzi}}, \bibinfo {author} {\bibfnamefont {A.}~\bibnamefont
  {G\"orlitz}}, \ and\ \bibinfo {author} {\bibfnamefont {T.}~\bibnamefont
  {Pfau}},\ }\href {\doibase 10.1103/PhysRevLett.89.130401} {\bibfield
  {journal} {\bibinfo  {journal} {Phys. Rev. Lett.}\ }\textbf {\bibinfo
  {volume} {89}},\ \bibinfo {pages} {130401} (\bibinfo {year}
  {2002})}\BibitemShut {NoStop}%
\bibitem [{\citenamefont {Pedri}\ and\ \citenamefont
  {Santos}(2005)}]{Pedri2005a}%
  \BibitemOpen
  \bibfield  {author} {\bibinfo {author} {\bibfnamefont {P.}~\bibnamefont
  {Pedri}}\ and\ \bibinfo {author} {\bibfnamefont {L.}~\bibnamefont {Santos}},\
  }\href {\doibase 10.1103/PhysRevLett.95.200404} {\bibfield  {journal}
  {\bibinfo  {journal} {Phys. Rev. Lett.}\ }\textbf {\bibinfo {volume} {95}},\
  \bibinfo {pages} {200404} (\bibinfo {year} {2005})}\BibitemShut {NoStop}%
\bibitem [{\citenamefont {Klawunn}\ and\ \citenamefont
  {Santos}(2009)}]{Klawunn2009a}%
  \BibitemOpen
  \bibfield  {author} {\bibinfo {author} {\bibfnamefont {M.}~\bibnamefont
  {Klawunn}}\ and\ \bibinfo {author} {\bibfnamefont {L.}~\bibnamefont
  {Santos}},\ }\href {\doibase 10.1103/PhysRevA.80.013611} {\bibfield
  {journal} {\bibinfo  {journal} {Phys. Rev. A}\ }\textbf {\bibinfo {volume}
  {80}},\ \bibinfo {pages} {013611} (\bibinfo {year} {2009})}\BibitemShut
  {NoStop}%
\bibitem [{\citenamefont {Junginger}\ \emph {et~al.}(2010)\citenamefont
  {Junginger}, \citenamefont {Main},\ and\ \citenamefont
  {Wunner}}]{Junginger2010a}%
  \BibitemOpen
  \bibfield  {author} {\bibinfo {author} {\bibfnamefont {A.}~\bibnamefont
  {Junginger}}, \bibinfo {author} {\bibfnamefont {J.}~\bibnamefont {Main}}, \
  and\ \bibinfo {author} {\bibfnamefont {G.}~\bibnamefont {Wunner}},\ }\href
  {\doibase 10.1103/PhysRevA.82.023602} {\bibfield  {journal} {\bibinfo
  {journal} {Phys. Rev. A}\ }\textbf {\bibinfo {volume} {82}},\ \bibinfo
  {pages} {023602} (\bibinfo {year} {2010})}\BibitemShut {NoStop}%
\bibitem [{\citenamefont {Ticknor}\ \emph {et~al.}(2011)\citenamefont
  {Ticknor}, \citenamefont {Wilson},\ and\ \citenamefont
  {Bohn}}]{Ticknor2011a}%
  \BibitemOpen
  \bibfield  {author} {\bibinfo {author} {\bibfnamefont {C.}~\bibnamefont
  {Ticknor}}, \bibinfo {author} {\bibfnamefont {R.~M.}\ \bibnamefont {Wilson}},
  \ and\ \bibinfo {author} {\bibfnamefont {J.~L.}\ \bibnamefont {Bohn}},\
  }\href {\doibase 10.1103/PhysRevLett.106.065301} {\bibfield  {journal}
  {\bibinfo  {journal} {Phys. Rev. Lett.}\ }\textbf {\bibinfo {volume} {106}},\
  \bibinfo {pages} {065301} (\bibinfo {year} {2011})}\BibitemShut {NoStop}%
\bibitem [{\citenamefont {Bismut}\ \emph {et~al.}(2012)\citenamefont {Bismut},
  \citenamefont {Laburthe-Tolra}, \citenamefont {Mar\'echal}, \citenamefont
  {Pedri}, \citenamefont {Gorceix},\ and\ \citenamefont
  {Vernac}}]{Bismut2012a}%
  \BibitemOpen
  \bibfield  {author} {\bibinfo {author} {\bibfnamefont {G.}~\bibnamefont
  {Bismut}}, \bibinfo {author} {\bibfnamefont {B.}~\bibnamefont
  {Laburthe-Tolra}}, \bibinfo {author} {\bibfnamefont {E.}~\bibnamefont
  {Mar\'echal}}, \bibinfo {author} {\bibfnamefont {P.}~\bibnamefont {Pedri}},
  \bibinfo {author} {\bibfnamefont {O.}~\bibnamefont {Gorceix}}, \ and\
  \bibinfo {author} {\bibfnamefont {L.}~\bibnamefont {Vernac}},\ }\href
  {\doibase 10.1103/PhysRevLett.109.155302} {\bibfield  {journal} {\bibinfo
  {journal} {Phys. Rev. Lett.}\ }\textbf {\bibinfo {volume} {109}},\ \bibinfo
  {pages} {155302} (\bibinfo {year} {2012})}\BibitemShut {NoStop}%
\bibitem [{\citenamefont {Baillie}\ \emph {et~al.}(2014)\citenamefont
  {Baillie}, \citenamefont {Bisset}, \citenamefont {Ticknor},\ and\
  \citenamefont {Blakie}}]{Baillie2014a}%
  \BibitemOpen
  \bibfield  {author} {\bibinfo {author} {\bibfnamefont {D.}~\bibnamefont
  {Baillie}}, \bibinfo {author} {\bibfnamefont {R.~N.}\ \bibnamefont {Bisset}},
  \bibinfo {author} {\bibfnamefont {C.}~\bibnamefont {Ticknor}}, \ and\
  \bibinfo {author} {\bibfnamefont {P.~B.}\ \bibnamefont {Blakie}},\ }\href
  {\doibase 10.1103/PhysRevLett.113.265301} {\bibfield  {journal} {\bibinfo
  {journal} {Phys. Rev. Lett.}\ }\textbf {\bibinfo {volume} {113}},\ \bibinfo
  {pages} {265301} (\bibinfo {year} {2014})}\BibitemShut {NoStop}%
\bibitem [{\citenamefont {Ronen}\ \emph {et~al.}(2006)\citenamefont {Ronen},
  \citenamefont {Bortolotti},\ and\ \citenamefont {Bohn}}]{Ronen2006a}%
  \BibitemOpen
  \bibfield  {author} {\bibinfo {author} {\bibfnamefont {S.}~\bibnamefont
  {Ronen}}, \bibinfo {author} {\bibfnamefont {D.~C.~E.}\ \bibnamefont
  {Bortolotti}}, \ and\ \bibinfo {author} {\bibfnamefont {J.~L.}\ \bibnamefont
  {Bohn}},\ }\href {\doibase 10.1103/PhysRevA.74.013623} {\bibfield  {journal}
  {\bibinfo  {journal} {Phys. Rev. A}\ }\textbf {\bibinfo {volume} {74}},\
  \bibinfo {pages} {013623} (\bibinfo {year} {2006})}\BibitemShut {NoStop}%
\bibitem [{\citenamefont {Fischer}(2006)}]{Fischer2006a}%
  \BibitemOpen
  \bibfield  {author} {\bibinfo {author} {\bibfnamefont {U.~R.}\ \bibnamefont
  {Fischer}},\ }\href {\doibase 10.1103/PhysRevA.73.031602} {\bibfield
  {journal} {\bibinfo  {journal} {Phys. Rev. A}\ }\textbf {\bibinfo {volume}
  {73}},\ \bibinfo {pages} {031602} (\bibinfo {year} {2006})}\BibitemShut
  {NoStop}%
\bibitem [{\citenamefont {Cai}\ \emph {et~al.}(2010)\citenamefont {Cai},
  \citenamefont {Rosenkranz}, \citenamefont {Lei},\ and\ \citenamefont
  {{Bao}}}]{Cai2010a}%
  \BibitemOpen
  \bibfield  {author} {\bibinfo {author} {\bibfnamefont {Y.}~\bibnamefont
  {Cai}}, \bibinfo {author} {\bibfnamefont {M.}~\bibnamefont {Rosenkranz}},
  \bibinfo {author} {\bibfnamefont {Z.}~\bibnamefont {Lei}}, \ and\ \bibinfo
  {author} {\bibfnamefont {W.}~\bibnamefont {{Bao}}},\ }\href {\doibase %
  10.1103/PhysRevA.82.043623} {\bibfield  {journal} {\bibinfo  {journal} {Phys.
  Rev. A}\ }\textbf {\bibinfo {volume} {82}},\ \bibinfo {pages} {043623}
  (\bibinfo {year} {2010})}\BibitemShut {NoStop}%
\bibitem [{\citenamefont {Parker}\ and\ \citenamefont
  {O'Dell}(2008)}]{Parker2008a}%
  \BibitemOpen
  \bibfield  {author} {\bibinfo {author} {\bibfnamefont {N.~G.}\ \bibnamefont
  {Parker}}\ and\ \bibinfo {author} {\bibfnamefont {D.~H.~J.}\ \bibnamefont
  {O'Dell}},\ }\href {\doibase 10.1103/PhysRevA.78.041601} {\bibfield
  {journal} {\bibinfo  {journal} {Phys. Rev. A}\ }\textbf {\bibinfo {volume}
  {78}},\ \bibinfo {pages} {041601} (\bibinfo {year} {2008})}\BibitemShut
  {NoStop}%
\bibitem [{\citenamefont {Wilson}\ and\ \citenamefont
  {Bohn}(2011)}]{Wilson2011a}%
  \BibitemOpen
  \bibfield  {author} {\bibinfo {author} {\bibfnamefont {R.~M.}\ \bibnamefont
  {Wilson}}\ and\ \bibinfo {author} {\bibfnamefont {J.~L.}\ \bibnamefont
  {Bohn}},\ }\href {\doibase 10.1103/PhysRevA.83.023623} {\bibfield  {journal}
  {\bibinfo  {journal} {Phys. Rev. A}\ }\textbf {\bibinfo {volume} {83}},\
  \bibinfo {pages} {023623} (\bibinfo {year} {2011})}\BibitemShut {NoStop}%
\bibitem [{\citenamefont {Wilson}\ \emph {et~al.}(2012)\citenamefont {Wilson},
  \citenamefont {Ticknor}, \citenamefont {Bohn},\ and\ \citenamefont
  {Timmermans}}]{Wilson2012a}%
  \BibitemOpen
  \bibfield  {author} {\bibinfo {author} {\bibfnamefont {R.~M.}\ \bibnamefont
  {Wilson}}, \bibinfo {author} {\bibfnamefont {C.}~\bibnamefont {Ticknor}},
  \bibinfo {author} {\bibfnamefont {J.~L.}\ \bibnamefont {Bohn}}, \ and\
  \bibinfo {author} {\bibfnamefont {E.}~\bibnamefont {Timmermans}},\ }\href
  {\doibase 10.1103/PhysRevA.86.033606} {\bibfield  {journal} {\bibinfo
  {journal} {Phys. Rev. A}\ }\textbf {\bibinfo {volume} {86}},\ \bibinfo
  {pages} {033606} (\bibinfo {year} {2012})}\BibitemShut {NoStop}%
\bibitem [{\citenamefont {Natu}\ and\ \citenamefont
  {Das~Sarma}(2013)}]{Natu2013a}%
  \BibitemOpen
  \bibfield  {author} {\bibinfo {author} {\bibfnamefont {S.~S.}\ \bibnamefont
  {Natu}}\ and\ \bibinfo {author} {\bibfnamefont {S.}~\bibnamefont
  {Das~Sarma}},\ }\href {\doibase 10.1103/PhysRevA.88.031604} {\bibfield
  {journal} {\bibinfo  {journal} {Phys. Rev. A}\ }\textbf {\bibinfo {volume}
  {88}},\ \bibinfo {pages} {031604} (\bibinfo {year} {2013})}\BibitemShut
  {NoStop}%
\bibitem [{\citenamefont {Natu}\ and\ \citenamefont
  {Wilson}(2013)}]{Natu2013b}%
  \BibitemOpen
  \bibfield  {author} {\bibinfo {author} {\bibfnamefont {S.~S.}\ \bibnamefont
  {Natu}}\ and\ \bibinfo {author} {\bibfnamefont {R.~M.}\ \bibnamefont
  {Wilson}},\ }\href {\doibase 10.1103/PhysRevA.88.063638} {\bibfield
  {journal} {\bibinfo  {journal} {Phys. Rev. A}\ }\textbf {\bibinfo {volume}
  {88}},\ \bibinfo {pages} {063638} (\bibinfo {year} {2013})}\BibitemShut
  {NoStop}%
\bibitem [{\citenamefont {Corson}\ \emph {et~al.}(2013)\citenamefont {Corson},
  \citenamefont {Wilson},\ and\ \citenamefont {Bohn}}]{Corson2013b}%
  \BibitemOpen
  \bibfield  {author} {\bibinfo {author} {\bibfnamefont {J.~P.}\ \bibnamefont
  {Corson}}, \bibinfo {author} {\bibfnamefont {R.~M.}\ \bibnamefont {Wilson}},
  \ and\ \bibinfo {author} {\bibfnamefont {J.~L.}\ \bibnamefont {Bohn}},\
  }\href {\doibase 10.1103/PhysRevA.88.013614} {\bibfield  {journal} {\bibinfo
  {journal} {Phys. Rev. A}\ }\textbf {\bibinfo {volume} {88}},\ \bibinfo
  {pages} {013614} (\bibinfo {year} {2013})}\BibitemShut {NoStop}%
\bibitem [{\citenamefont {Mulkerin}\ \emph {et~al.}(2013)\citenamefont
  {Mulkerin}, \citenamefont {van Bijnen}, \citenamefont {O'Dell}, \citenamefont
  {Martin},\ and\ \citenamefont {Parker}}]{Mulkerin2013a}%
  \BibitemOpen
  \bibfield  {author} {\bibinfo {author} {\bibfnamefont {B.~C.}\ \bibnamefont
  {Mulkerin}}, \bibinfo {author} {\bibfnamefont {R.~M.~W.}\ \bibnamefont {van
  Bijnen}}, \bibinfo {author} {\bibfnamefont {D.~H.~J.}\ \bibnamefont
  {O'Dell}}, \bibinfo {author} {\bibfnamefont {A.~M.}\ \bibnamefont {Martin}},
  \ and\ \bibinfo {author} {\bibfnamefont {N.~G.}\ \bibnamefont {Parker}},\
  }\href {\doibase 10.1103/PhysRevLett.111.170402} {\bibfield  {journal}
  {\bibinfo  {journal} {Phys. Rev. Lett.}\ }\textbf {\bibinfo {volume} {111}},\
  \bibinfo {pages} {170402} (\bibinfo {year} {2013})}\BibitemShut {NoStop}%
\bibitem [{\citenamefont {Yi}\ and\ \citenamefont {You}(2000)}]{Yi2000a}%
  \BibitemOpen
  \bibfield  {author} {\bibinfo {author} {\bibfnamefont {S.}~\bibnamefont
  {Yi}}\ and\ \bibinfo {author} {\bibfnamefont {L.}~\bibnamefont {You}},\
  }\href {\doibase 10.1103/PhysRevA.61.041604} {\bibfield  {journal} {\bibinfo
  {journal} {Phys. Rev. A}\ }\textbf {\bibinfo {volume} {61}},\ \bibinfo
  {pages} {041604} (\bibinfo {year} {2000})}\BibitemShut {NoStop}%
\bibitem [{\citenamefont {Yi}\ and\ \citenamefont {You}(2001)}]{Yi2001a}%
  \BibitemOpen
  \bibfield  {author} {\bibinfo {author} {\bibfnamefont {S.}~\bibnamefont
  {Yi}}\ and\ \bibinfo {author} {\bibfnamefont {L.}~\bibnamefont {You}},\
  }\href {\doibase 10.1103/PhysRevA.63.053607} {\bibfield  {journal} {\bibinfo
  {journal} {Phys. Rev. A}\ }\textbf {\bibinfo {volume} {63}},\ \bibinfo
  {pages} {053607} (\bibinfo {year} {2001})}\BibitemShut {NoStop}%
\bibitem [{\citenamefont {Modugno}\ \emph {et~al.}(2003)\citenamefont
  {Modugno}, \citenamefont {Pricoupenko},\ and\ \citenamefont
  {Castin}}]{Modugno2002a}%
  \BibitemOpen
  \bibfield  {author} {\bibinfo {author} {\bibfnamefont {M.}~\bibnamefont
  {Modugno}}, \bibinfo {author} {\bibfnamefont {L.}~\bibnamefont
  {Pricoupenko}}, \ and\ \bibinfo {author} {\bibfnamefont {Y.}~\bibnamefont
  {Castin}},\ }\href {\doibase 10.1140/epjd/e2003-00015-y} {\bibfield
  {journal} {\bibinfo  {journal} {Eur. Phys. J. D}\ }\textbf {\bibinfo {volume}
  {22}},\ \bibinfo {pages} {235} (\bibinfo {year} {2003})}\BibitemShut
  {NoStop}%
\bibitem [{\citenamefont {Kelly}(2003)}]{KellyBook}%
  \BibitemOpen
  \bibfield  {author} {\bibinfo {author} {\bibfnamefont {C.~T.}\ \bibnamefont
  {Kelly}},\ }\href@noop {} {\emph {\bibinfo {title} {{Solving Nonlinear
  Equations with Newton's Method}}}},\ Fundamentals of Algorithms\ (\bibinfo
  {publisher} {SIAM},\ \bibinfo {address} {Philadelphia},\ \bibinfo {year}
  {2003})\BibitemShut {NoStop}%
\bibitem [{\citenamefont {Griffin}(1996)}]{Griffin1996a}%
  \BibitemOpen
  \bibfield  {author} {\bibinfo {author} {\bibfnamefont {A.}~\bibnamefont
  {Griffin}},\ }\href {\doibase 10.1103/PhysRevB.53.9341} {\bibfield  {journal}
  {\bibinfo  {journal} {Phys. Rev. B}\ }\textbf {\bibinfo {volume} {53}},\
  \bibinfo {pages} {9341} (\bibinfo {year} {1996})}\BibitemShut {NoStop}%
\bibitem [{\citenamefont {Morgan}\ \emph {et~al.}(1998)\citenamefont {Morgan},
  \citenamefont {Choi}, \citenamefont {Burnett},\ and\ \citenamefont
  {Edwards}}]{Morgan1998a}%
  \BibitemOpen
  \bibfield  {author} {\bibinfo {author} {\bibfnamefont {S.~A.}\ \bibnamefont
  {Morgan}}, \bibinfo {author} {\bibfnamefont {S.}~\bibnamefont {Choi}},
  \bibinfo {author} {\bibfnamefont {K.}~\bibnamefont {Burnett}}, \ and\
  \bibinfo {author} {\bibfnamefont {M.}~\bibnamefont {Edwards}},\ }\href
  {\doibase 10.1103/PhysRevA.57.3818} {\bibfield  {journal} {\bibinfo
  {journal} {Phys. Rev. A}\ }\textbf {\bibinfo {volume} {57}},\ \bibinfo
  {pages} {3818} (\bibinfo {year} {1998})}\BibitemShut {NoStop}%
\bibitem [{\citenamefont {Garbow}(1974)}]{Garbow1974a}%
  \BibitemOpen
  \bibfield  {author} {\bibinfo {author} {\bibfnamefont {B.~S.}\ \bibnamefont
  {Garbow}},\ }\href {\doibase 10.1016/0010-4655(74)90086-1} {\bibfield
  {journal} {\bibinfo  {journal} {Comput. Phys. Commun.}\ }\textbf {\bibinfo
  {volume} {7}},\ \bibinfo {pages} {179 } (\bibinfo {year} {1974})}\BibitemShut
  {NoStop}%
\bibitem [{\citenamefont {Wilson}\ \emph {et~al.}(2009)\citenamefont {Wilson},
  \citenamefont {Ronen},\ and\ \citenamefont {Bohn}}]{Wilson2009a}%
  \BibitemOpen
  \bibfield  {author} {\bibinfo {author} {\bibfnamefont {R.~M.}\ \bibnamefont
  {Wilson}}, \bibinfo {author} {\bibfnamefont {S.}~\bibnamefont {Ronen}}, \
  and\ \bibinfo {author} {\bibfnamefont {J.~L.}\ \bibnamefont {Bohn}},\ }\href
  {\doibase 10.1103/PhysRevA.80.023614} {\bibfield  {journal} {\bibinfo
  {journal} {Phys. Rev. A}\ }\textbf {\bibinfo {volume} {80}},\ \bibinfo
  {pages} {023614} (\bibinfo {year} {2009})}\BibitemShut {NoStop}%
\bibitem [{\citenamefont {Bohn}\ \emph {et~al.}(2009)\citenamefont {Bohn},
  \citenamefont {Wilson},\ and\ \citenamefont {Ronen}}]{Bohn2009a}%
  \BibitemOpen
  \bibfield  {author} {\bibinfo {author} {\bibfnamefont {J.}~\bibnamefont
  {Bohn}}, \bibinfo {author} {\bibfnamefont {R.}~\bibnamefont {Wilson}}, \ and\
  \bibinfo {author} {\bibfnamefont {S.}~\bibnamefont {Ronen}},\ }\href
  {\doibase 10.1134/S1054660X09040021} {\bibfield  {journal} {\bibinfo
  {journal} {Laser Phys.}\ }\textbf {\bibinfo {volume} {19}},\ \bibinfo {pages}
  {547} (\bibinfo {year} {2009})}\BibitemShut {NoStop}%
\bibitem [{\citenamefont {Parker}\ \emph {et~al.}(2009)\citenamefont {Parker},
  \citenamefont {Ticknor}, \citenamefont {Martin},\ and\ \citenamefont
  {O'Dell}}]{Parker2009a}%
  \BibitemOpen
  \bibfield  {author} {\bibinfo {author} {\bibfnamefont {N.~G.}\ \bibnamefont
  {Parker}}, \bibinfo {author} {\bibfnamefont {C.}~\bibnamefont {Ticknor}},
  \bibinfo {author} {\bibfnamefont {A.~M.}\ \bibnamefont {Martin}}, \ and\
  \bibinfo {author} {\bibfnamefont {D.~H.~J.}\ \bibnamefont {O'Dell}},\ }\href
  {\doibase 10.1103/PhysRevA.79.013617} {\bibfield  {journal} {\bibinfo
  {journal} {Phys. Rev. A}\ }\textbf {\bibinfo {volume} {79}},\ \bibinfo {eid}
  {013617} (\bibinfo {year} {2009})}\BibitemShut {NoStop}%
\bibitem [{\citenamefont {Billy}\ \emph {et~al.}(2012)\citenamefont {Billy},
  \citenamefont {Henn}, \citenamefont {M\"uller}, \citenamefont {Maier},
  \citenamefont {Kadau}, \citenamefont {Griesmaier}, \citenamefont
  {Jona-Lasinio}, \citenamefont {Santos},\ and\ \citenamefont
  {Pfau}}]{Billy2012a}%
  \BibitemOpen
  \bibfield  {author} {\bibinfo {author} {\bibfnamefont {J.}~\bibnamefont
  {Billy}}, \bibinfo {author} {\bibfnamefont {E.~A.~L.}\ \bibnamefont {Henn}},
  \bibinfo {author} {\bibfnamefont {S.}~\bibnamefont {M\"uller}}, \bibinfo
  {author} {\bibfnamefont {T.}~\bibnamefont {Maier}}, \bibinfo {author}
  {\bibfnamefont {H.}~\bibnamefont {Kadau}}, \bibinfo {author} {\bibfnamefont
  {A.}~\bibnamefont {Griesmaier}}, \bibinfo {author} {\bibfnamefont
  {M.}~\bibnamefont {Jona-Lasinio}}, \bibinfo {author} {\bibfnamefont
  {L.}~\bibnamefont {Santos}}, \ and\ \bibinfo {author} {\bibfnamefont
  {T.}~\bibnamefont {Pfau}},\ }\href {\doibase 10.1103/PhysRevA.86.051603}
  {\bibfield  {journal} {\bibinfo  {journal} {Phys. Rev. A}\ }\textbf {\bibinfo
  {volume} {86}},\ \bibinfo {pages} {051603} (\bibinfo {year}
  {2012})}\BibitemShut {NoStop}%
\bibitem [{\citenamefont {Dutta}\ and\ \citenamefont
  {Meystre}(2007)}]{Dutta2007a}%
  \BibitemOpen
  \bibfield  {author} {\bibinfo {author} {\bibfnamefont {O.}~\bibnamefont
  {Dutta}}\ and\ \bibinfo {author} {\bibfnamefont {P.}~\bibnamefont
  {Meystre}},\ }\href {\doibase 10.1103/PhysRevA.75.053604} {\bibfield
  {journal} {\bibinfo  {journal} {Phys. Rev. A}\ }\textbf {\bibinfo {volume}
  {75}},\ \bibinfo {pages} {053604} (\bibinfo {year} {2007})}\BibitemShut
  {NoStop}%
\bibitem [{\citenamefont {Linscott}\ and\ \citenamefont
  {Blakie}(2014)}]{Linscott2014a}%
  \BibitemOpen
  \bibfield  {author} {\bibinfo {author} {\bibfnamefont {E.~B.}\ \bibnamefont
  {Linscott}}\ and\ \bibinfo {author} {\bibfnamefont {P.~B.}\ \bibnamefont
  {Blakie}},\ }\href {\doibase 10.1103/PhysRevA.90.053605} {\bibfield
  {journal} {\bibinfo  {journal} {Phys. Rev. A}\ }\textbf {\bibinfo {volume}
  {90}},\ \bibinfo {pages} {053605} (\bibinfo {year} {2014})}\BibitemShut
  {NoStop}%
\bibitem [{\citenamefont {Edwards}\ \emph {et~al.}(2012)\citenamefont
  {Edwards}, \citenamefont {Krygier}, \citenamefont {Seddiqi}, \citenamefont
  {Benton},\ and\ \citenamefont {Clark}}]{Edwards2012a}%
  \BibitemOpen
  \bibfield  {author} {\bibinfo {author} {\bibfnamefont {M.}~\bibnamefont
  {Edwards}}, \bibinfo {author} {\bibfnamefont {M.}~\bibnamefont {Krygier}},
  \bibinfo {author} {\bibfnamefont {H.}~\bibnamefont {Seddiqi}}, \bibinfo
  {author} {\bibfnamefont {B.}~\bibnamefont {Benton}}, \ and\ \bibinfo {author}
  {\bibfnamefont {C.}~\bibnamefont {Clark}},\ }\href {\doibase %
  10.1103/PhysRevE.86.056710} {\bibfield  {journal} {\bibinfo  {journal} {Phys.
  Rev. E}\ }\textbf {\bibinfo {volume} {86}},\ \bibinfo {pages} {056710}
  (\bibinfo {year} {2012})}\BibitemShut {NoStop}%
\bibitem [{\citenamefont {Sykes}\ and\ \citenamefont {Ticknor}()}]{Sykes2012a}%
  \BibitemOpen
  \bibfield  {author} {\bibinfo {author} {\bibfnamefont {A.~G.}\ \bibnamefont
  {Sykes}}\ and\ \bibinfo {author} {\bibfnamefont {C.}~\bibnamefont
  {Ticknor}},\ }\href@noop {} {\ }\Eprint {http://arxiv.org/abs/1206.1350}
  {arXiv:1206.1350} \BibitemShut {NoStop}%
\bibitem [{\citenamefont {Ronen}\ and\ \citenamefont
  {Bohn}(2007)}]{Ronen2007b}%
  \BibitemOpen
  \bibfield  {author} {\bibinfo {author} {\bibfnamefont {S.}~\bibnamefont
  {Ronen}}\ and\ \bibinfo {author} {\bibfnamefont {J.~L.}\ \bibnamefont
  {Bohn}},\ }\href {\doibase 10.1103/PhysRevA.76.043607} {\bibfield  {journal}
  {\bibinfo  {journal} {Phys. Rev. A}\ }\textbf {\bibinfo {volume} {76}},\
  \bibinfo {pages} {043607} (\bibinfo {year} {2007})}\BibitemShut {NoStop}%
\bibitem [{\citenamefont {Bisset}\ \emph {et~al.}(2012)\citenamefont {Bisset},
  \citenamefont {Baillie},\ and\ \citenamefont {Blakie}}]{Bisset2012}%
  \BibitemOpen
  \bibfield  {author} {\bibinfo {author} {\bibfnamefont {R.~N.}\ \bibnamefont
  {Bisset}}, \bibinfo {author} {\bibfnamefont {D.}~\bibnamefont {Baillie}}, \
  and\ \bibinfo {author} {\bibfnamefont {P.~B.}\ \bibnamefont {Blakie}},\
  }\href {\doibase 10.1103/PhysRevA.86.033609} {\bibfield  {journal} {\bibinfo
  {journal} {Phys. Rev. A}\ }\textbf {\bibinfo {volume} {86}},\ \bibinfo
  {pages} {033609} (\bibinfo {year} {2012})}\BibitemShut {NoStop}%
\bibitem [{\citenamefont {Ticknor}(2012)}]{Ticknor2012a}%
  \BibitemOpen
  \bibfield  {author} {\bibinfo {author} {\bibfnamefont {C.}~\bibnamefont
  {Ticknor}},\ }\href {\doibase 10.1103/PhysRevA.85.033629} {\bibfield
  {journal} {\bibinfo  {journal} {Phys. Rev. A}\ }\textbf {\bibinfo {volume}
  {85}},\ \bibinfo {pages} {033629} (\bibinfo {year} {2012})}\BibitemShut
  {NoStop}%
\end{thebibliography}

%

\end{document}